\documentclass{emulateapj}
\usepackage{graphicx,natbib,psfig,textcomp,fontenc}

\shorttitle{The 1.6 $\mu$m near infrared nuclei of 3CR galaxies}
\shortauthors{Baldi R.~D. et al.}
\begin{document}

\title{The 1.6 micron near infrared nuclei of 3C radio galaxies: \\
Jets, thermal emission or scattered light?$^{*}$}\thanks{$^{*}$ Based on
observations made with the NASA/ESA Hubble Space Telescope, obtained at the
Space Telescope Science Institute, which is operated by the Association of
Universities for Research in Astronomy, Inc..}

\author{Ranieri D. Baldi\altaffilmark{1,2}, Marco Chiaberge\altaffilmark{2,3},
Alessandro Capetti\altaffilmark{4}, William Sparks\altaffilmark{2}, F. Duccio
Macchetto\altaffilmark{2}, Christopher P. O'Dea\altaffilmark{5}, David
J. Axon\altaffilmark{5}, Stefi A. Baum\altaffilmark{5}, and Alice
C. Quillen\altaffilmark{6}}

\email{baldi@oato.inaf.it}

\altaffiltext{1}{Universit\'{a} di Torino, via P. Giuria 1, I-10125 Torino, Italy; baldi@oato.inaf.it}
\altaffiltext{2}{Space Telescope Science Institute, 3700 San Martin Drive,
  Baltimore, Baltimore, MD 21218}
\altaffiltext{3}{INAF-Istituto di Radio
  Astronomia, via P. Gobetti 101, I-40129 Bologna, Italy}
\altaffiltext{4}{INAF-Osservatorio Astronomico di Torino, Strada Osservatorio 20,
  10025 Pino Torinese, Italy}
 \altaffiltext{5}{Department of Physics, Rochester Institute of Technology,
   Carlson Center for Imaging Science 76-3144, 84 Lomb Memorial Dr.,
   Rochester, NY 14623, USA}
\altaffiltext{6}{Department of Physics and Astronomy, University of Rochester,
  Rochester, NY 14627, USA}

\begin{abstract}

Using HST NICMOS 2 observations we have measured 1.6 $\mu$m near infrared
nuclear luminosities of 100 3CR radio galaxies with z $<$ 0.3, by modeling and
subtracting the extended emission from the host galaxy. We performed a
multi-wavelength statistical analysis (including optical and radio data) of
the properties of the nuclei following classification of the objects into FR~I and
FR~II, and LIG (low-ionization galaxies), HIG (high-ionization galaxies) and
BLO (broad-lined objects) using the radio morphology and optical spectra,
respectively.  The correlations among near infrared, optical, and radio
nuclear luminosity support the idea that the near infrared nuclear emission of
FR~Is has a non-thermal origin. Despite the difference in radio morphology, the
multi-wavelength properties of FR~II LIG nuclei are statistically
indistinguishable from those of FR~Is, an indication of a common structure of
the central engine. All BLOs show an unresolved near infrared nucleus and a
large near infrared excess with respect to FR~II LIGs and FR~Is of equal radio
core luminosity. This requires the presence of an additional (and dominant)
component other than the non-thermal light. Considering the shape of their
spectral energy distribution, we ascribe the origin of their near infrared
light to hot circumnuclear dust. A near infrared excess is also found in HIGs,
but their nuclei are substantially fainter than those of BLO.  This result
indicates that substantial obscuration along the line-of-sight to the nuclei
is still present at 1.6 $\mu$m.  Nonetheless, HIGs nuclei cannot simply be
explained in terms of dust obscuration: a significant contribution from light
reflected in a circumnuclear scattering region is needed to account for their
multiwavelength properties.
\end{abstract}


\section{Introduction}

Infrared observations of radio galaxies are a useful tool to explore the
physics of their nuclei in the framework of the Unified Models for Active
Galactic Nuclei (AGN) as in this band the impact of dust absorption is
strongly reduced with respect to the optical.  In particular, the study of the
infrared nuclear sources allows us to investigate the `scheme' unifying
different classes of AGN, by exploring the properties of their accretion disks
and the presence of any absorbing material that can account for the complex
AGN taxonomy.

In the ``zeroth-order approximation'' of the AGN unification scheme for
radio-loud sources (e.g. \citealt{urry95}), powerful radio galaxies with FR~II
edge-brightened morphology \citep{fanaroff74} are believed to be misaligned
quasars, while lower power, edge-darkened FR~Is are associated with BL~Lac
objects. However, it is clear that this zeroth order approximation picture
based on statistical comparison of the properties of radio-loud AGN is
over-simplified. Inconsistencies between the characteristics of the ``parent''
and ``beamed'' populations are well reported in the literature: for example,
several BL~Lacs show broad emission lines (e.g. \citealt{vermeulen95}); and
there are inconsistent environments (e.g. \citealt{zirbel97,owen96}) and
inconsistent radio morphologies (e.g. \citealt{antonucci86}) (see also
\citealt{urry95} and \citealt{chiaberge04} for reviews on this subject).

Furthermore, how the large-scale radio properties relate to different nuclear
properties is still not clear. In fact, there is mixed evidence for a
correlation between the morphological FR~I/FR~II dichotomy and different
levels of nuclear activity
\citep{baum95,chiaberge99,xu99b,chiaberge02,gijs02}. Another classification,
based on the optical narrow emission lines ratios
\citep{hine79,laing94,jackson97,buttiglione10} introduces two groups of
radio-loud AGN: the Low Excitation Galaxies (LEG) and High Excitation Galaxies
(HEG)\footnote{In the following we will refer to those objects as High
Ionization Galaxies (HIG) and Low Ionization Galaxies (LIG). See Sect.~2 for
details.}. Radio galaxies with FR~I radio morphology almost always belong to
LEG class, while nearly all HEGs are FR~II radio sources. However, there is a
group of FR~II LEGs whose place in the unifying scenario is still largely not
understood. \citet{wall97} and \citet{jackson99} proposed that such objects
constitute a single population of radio galaxies together with FR~Is. In line
with this idea, the two spectroscopic classes, LEG and HEG, show significant
differences in several properties: star formation
(e.g. \citealt{baldi08,smolcic09b}), environment
(e.g. \citealt{chiaberge00,hardcastle04}), triggering mechanism of nuclear
activity (gas-rich or gas-poor merger) (e.g. \citealt{baldi08}); there are
indications that they differ also in terms of accretion rate and radiative
efficiency of disk (e.g. \citealt{marchesini04,balmaverde08}); contrarily to
what is seen in HEG, LEG do not show evidence for large absorbing column
densities in the X-ray spectra (e.g. \citealt{hardcastle06}), for warm dusty
tori in mid-IR images \citep{vanderwolk09}, and for broad lines in their
optical spectra (e.g. \citealt{buttiglione10}). Taken together these
differences suggest that high and low excitation galaxies belong to different
classes of radio-loud nuclear activity which are associated with distinct
environmental conditions. Furthermore, it has been suggested that in the
framework of the unification model, HEG and LEG might represent the parent
population of QSOs and BL~Lacs respectively (e.g. \citealt{jackson99}).

A number of recent papers in the literature address the topic of the infrared
view of the active nucleus of radio galaxies. \citet{dicken09} show that the
infrared emission at both 24 and 70 micron can be explained in terms of dust
heated predominantly by the AGN.  More comprehensive mid-IR and far-IR
spectrophotometry for the z $<$ 1 3CR objects has been obtained from ISO
(\citealt{siebenmorgen04,haas04}) as well as from Spitzer (e.g.,
\citealt{shi05,haas05,ogle06,cleary07,leipski09}). For high-redshift 3CR
sources near infrared spectrophotometry using Spitzer has been presented by
\citet{haas08} and \citet{netzer04}.

The mid-infrared observations are sensitive to thermal emission from warm dust
and have revealed rather complex properties within the different radio-galaxy
classifications. The majority of FRIs are faint at those wavelengths, implying
that the bulk kinetic power of the radio jets is the energetically dominant
mechanism in most of these sources \citep{whysong04}.  Even in FR~Is this is
not a universal conclusion as \citet{haas04} with ISO data and
\citet{leipski09} with Spitzer data show that in some FR~Is a dust emission
bump is detected. On the other hand, \citet{ogle06} found that about 50\% of
the FR~II narrow-line radio galaxies have mid-infrared photometric and
spectroscopic properties comparable to quasars of matched radio luminosity,
while there is a subsample of mid-IR weak FR~II radio galaxies which may
constitute a separate population of non-thermal, jet-dominated sources with
low accretion power. \citet{haas04} present evidence that the MIR data do in
fact support the orientation-dependent unification scheme of the powerful
FR~II galaxies with quasars. This has been subsequently confirmed by further
work, e.g. \citet{haas08,ogle06,dicken09}.

This paper has the aim of investigating AGN activity by focusing on the
nuclear 1.6 $\mu$m near infrared emission, identifying its physical origin and
relating it to both the radio morphological and optical spectroscopic classes.
\citet{chiaberge99,chiaberge02} studied the HST optical images of the nuclear
regions of radio galaxies belonging to 3CR sample with the same purpose. The
NIR snapshot survey of 3CR objects with $z<0.3$ provided us with a larger
coverage of the sample than that in the optical band.  Furthermore, the
fraction of 3CR sources with spectroscopic classification is now substantially
larger \citep{buttiglione10}, allowing us to explore in greater detail the
behavior of the nuclei of each class.

The resolution of HST images is suitable for this purpose
and allow us to distinguish the NIR emission of stellar host galaxy
background from the genuine AGN radiation.

The structure of the paper is as follows. In Sect.~2 we describe the sample
and in Sect.~3 we derive the surface brightness profiles which we fit with
either S\'{e}rsic or core-S\'{e}rsic models. The multiwavelength properties of
our sample are discussed in Sect.~4 in order to identify the origin of the NIR
nuclear emission tackled in Sect.~5. In Sect.~6 we summarize and draw
conclusions. In Appendix we compare the FR~I radio core flux observed
with VLA and VLBI.

We adopt a Hubble constant of H$_{0}$ = 75 km s$^{-1}$ Mpc$^{-1}$ and q$_{0}$
= 0.5.  We used this cosmology in order to be in accordance with the similar
results obtained by \citet{chiaberge99}, \citet{chiaberge00},
\citet{chiaberge02}.  Assuming a different cosmology would not affect our
results since the redshift of the sources considered in this paper is limited
to $z=0.3$.

\section{Sample and HST observations}

The sample of galaxies we analyzed belongs to the Revised Third Cambridge
Catalog 3CR (\citealt{bennett62a}, \citealt{bennett62b}, \citealt{spinrad85}).
Since it is selected based on the low-frequency radio flux (S$_{178 MHz}$ $>$
10 Jy), the sample is essentially free from orientation bias. We chose all 116
3CR sources with z $<$ 0.3 with the primary goal of characterizing the radio
galaxy hosts mostly free from the effects of dust.  One hundred objects were
observed as part of HST/NICMOS 3CR snapshot program GO~10173 (PI: Sparks).
The NIR images are published in \citet{madrid06} and \citet{floyd08}. Only
eleven galaxies were observed as a part of other programs and details of these
are given in Table~\ref{log}. We also included
in the sample the FR~I radio galaxy NGC~6251 because, as pointed out by
\citet{waggett77} it should be included in 3CR catalog.

All of the objects of our sample were observed with HST NICMOS Camera 2
(NIC2), which has a field of view of 19$\farcs$2 $\times$ 19$\farcs$2 and a
projected pixel size of 0.075\arcsec, using the F160W filter (similar to H
band), which is centered at 1.60 $\mu$m, covering a wavelength range from 1.4
to 1.8 $\mu$m and which includes the Pa$\beta$ line for objects with z $>$
0.1. Using the broad and narrow line H$\alpha$ fluxes measured for
3CR galaxies \citep{buttiglione09} over a 2$\arcsec \times 2\arcsec$ aperture,
we have estimate the contamination from line emission in the H band. In the
case of the narrow lines we assumed ratio Pa$\beta$/H$\alpha$ = 0.06 from
atomic physics \citep{osterbrock89}. Since the broad line galaxies are know
not to conform to case B we have used the observed near infrared broad line
region ratios observed in type 1 AGN by \citet{riffel06} and \citet{landt08}
to estimate the contamination in the F160W filter.  This analysis shows the
contamination to be $\lesssim$ 15\% for Broad Line Objects and $\lesssim$ 4\%
for Narrow Line Objects and in both cases that is small enough not to affect
our conclusions regarding the origin of the nuclear NIR emission.

Due to technical difficulties, 3C~410 and 3C~442 were only observed for 575.9
s, while all the remaining galaxies were observed for a total exposure time of
1151.8 s. A detailed description of the observations and data reduction can be
found in \citet{madrid06}.

\begin{table}
\begin{center}
\caption{Log of observations}
\begin{tabular}{ccc}
\tableline\tableline
Name & Date & t$_{exp}$ \\
     & yy-mm-dd &  s       \\
\tableline
3C~084    &  1998-03-16 & 639.9   \\
3C~264    &  1998-05-12 & 447.7   \\ 
3C~270    &  1998-04-23	& 767.8   \\
3C~272.1  &  1998-07-13	& 767.8   \\ 
3C~274    &  1997-11-20	& 127.9   \\
3C~293    &  1998-08-19	& 2751.8  \\ 
3C~305    &  1998-07-19	& 2783.8  \\
3C~317    &  1998-08-26	& 639.8   \\
3C~338    &  1997-12-18	& 4159.8  \\
3C~405    &  1997-12-16	& 1343.9  \\
NGC~6251  &  1998-07-06	& 543.5   \\
\tableline
\end{tabular}
\label{log}
\tablecomments{Log of observations for the objects not included in the HST
  1.6 $\mu$m near infrared snapshot program 10173 (PI: Sparks)
  (\citealt{madrid06,floyd08}). All data are taken with the same
  configuration with Camera NIC2 and F160W filter.}
\end{center}
\end{table}

Each galaxy was classified in two ways. Firstly, into FR~I and FR~II based on
radio morphology \citep{fanaroff74}: the radio morphological classification,
mostly being adopted from \citet{zirbel95}. Secondly, based on the optical
broad line strengths and narrow emission lines ratios (from
\citealt{buttiglione09, buttiglione10}) we classified them into the stand
groupings of High and Low Ionization Galaxies (HIG and LIG) and Broad Line
Objects (BLO). Note that the BLO objects belong to HIG class because they
show high-ionization narrow emission lines. We adopt the HIG/LIG nomenclature
to better represent the physical condition of the narrow line region gas in
these objects. This classification is however entirely consistent with the
HEG/LEG scheme widely adopted in the literature (e.g.,
\citealt{hine79,laing94,buttiglione10}).  Following \citet{buttiglione10} we
introduced a 4th class of Extremely-Low Excitation Galaxies (ELEG),
characterized by an extremely low value of [O~III]/H$\beta$ $\sim$ 0.5,
$\sim$6 times lower than usually observed in 3CR/LIG. In this work we
consider ELEGs as a separate class of objects.  The objects for which we do
not have new spectral data are classified according to \citet{jackson97}.  In
the sample there is also one galaxy, 3C~198, characterized by a star forming
spectrum (marked as ``SF'' in Table~\ref{table1}) and a BL~Lac object (3C~371)
which are not considered in the following statistical evaluations because of
their unique properties among the objects of our sample. For one galaxy
(3C~410) we could not find any reliable spectral classification in the
literature.

\section{Radial brightness profile fitting and nuclear luminosity measuring}
\label{method}

\begin{figure*}
\includegraphics[scale=0.7]{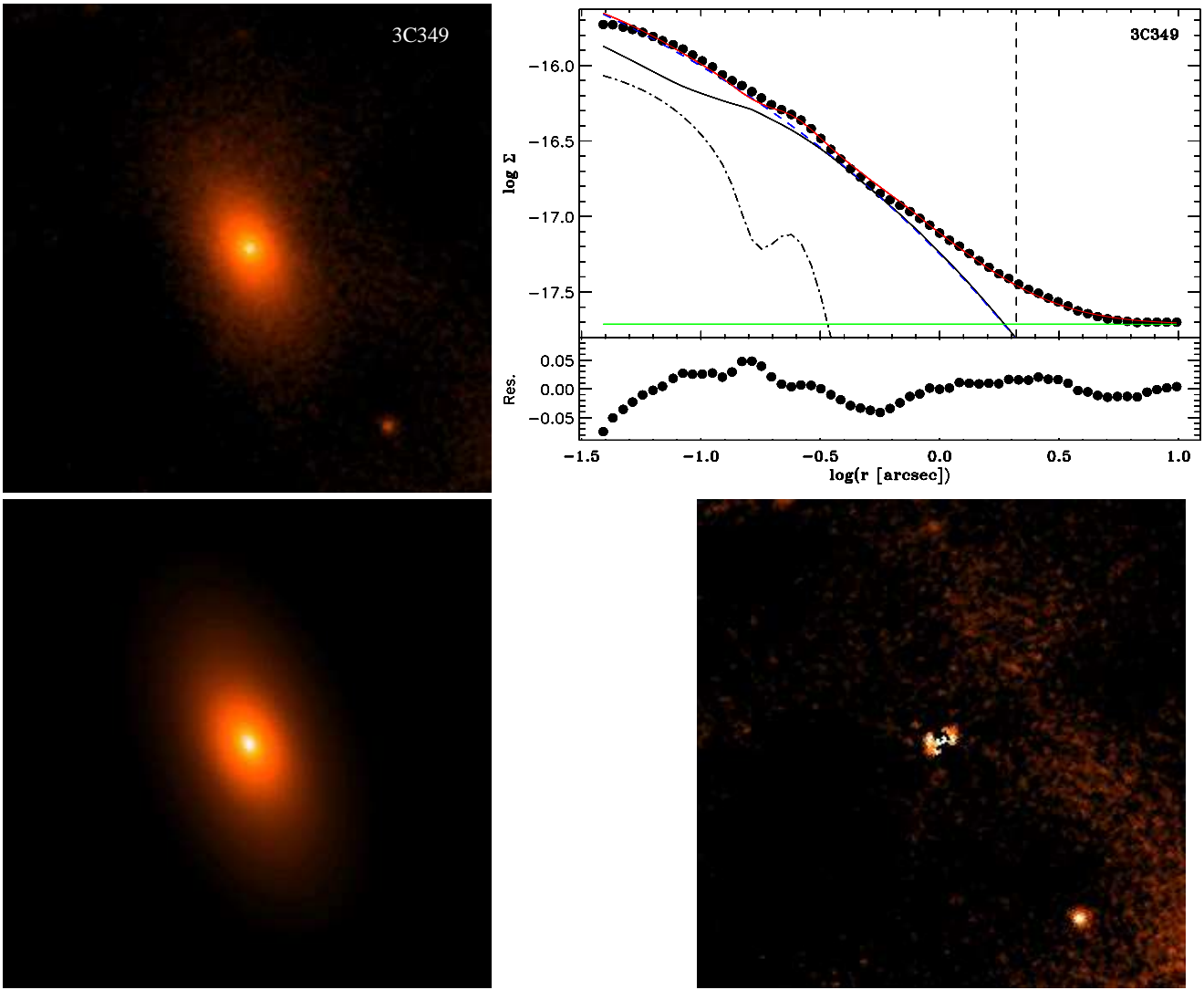}
\caption{This figure summarizes the fitting procedure of the surface
brightness profile (in this example 3C~349). In the upper left panel we show
the original NICMOS image. The lower right panel presents the model image
using the parameters obtained by the task ellipse. The lower left panel shows
the residuals image (typically $\sim$10\% in the nuclear region). The object
in the S-W corner in the residual image was masked and not fitted by the
ellipse task. All the images are $\sim$7\arcsec\ wide. The upper right panel
shows the fitting of the radial unidimensional brightness profile (in units of
erg s$^{-1}$ cm$^{-2}$ \AA$^{-1}$ arcsec$^{-2}$) with a S\'{e}rsic law (blue
dashed curve), a nuclear component (black dot-dashed line) and a background
level (green horizontal line). We take into account in the fit the convolution
(black solid line) of the model with the instrumental PSF. The final model is
the sum of each component and is the red solid line. The vertical dashed line
gives the location of the effective radius r$_{e}$. In the lower region of the
panel we show the residuals.}
\label{esempio}
\end{figure*}

The aim of this paper is to measure the NIR nuclear sources. In order to
disentangle such a component from the underlying stellar emission of the host
galaxy, it is necessary to model the distribution of the galactic light. To
this purpose we derived the surface brightness profile of all our
sources. Radial profiles can be derived using a number of different methods:
ellipse fitting to the isophotes of the galaxies (e.g., \citealt{heraudeau96})
obtaining a one-dimensional analysis (e.g., \citealt{baggett98}), and full
two-dimensional analysis (e.g., \citealt{byun95,peng02}). In this paper, we
adopted the first approach.

The ``ellipse'' task in IRAF STSDAS \citep{jedrzejewski87} is suitable to
analyzing the axisymmetric structure, evident in most objects. This task
allows us to trace the radial profile and radial changes in ellipticity,
position angle, and coordinates of the center of the isophotes.

Most of the host galaxies of 3CR sources appear as smooth ellipticals in
infrared. Only in few cases are distortions due to dust lanes, jets, or
plumes.  Before deriving the galaxy profile, we mask out all spurious objects
such as stars, jets, dust lanes, bright, galaxy companions and residual
diffraction patterns associated with bright stars. When dust features are
present in the nuclear region we adopted a multi-stage strategy. To stabilize
the fit in the radial range affected by the dust we started by fixing some or
all fit-parameters to the values of the last isophote that is not affected by
the dust. We then produce an initial model image and we subtracted it from the
original data to unveil the presence of further faint sources and/or dust
features, that were then also masked out and so on until no further feature
were found.

In Fig.~\ref{esempio} we show the stages
of the procedure to derive and to fit the radial brightness
profile.

\begin{figure*}
\includegraphics[scale=0.65]{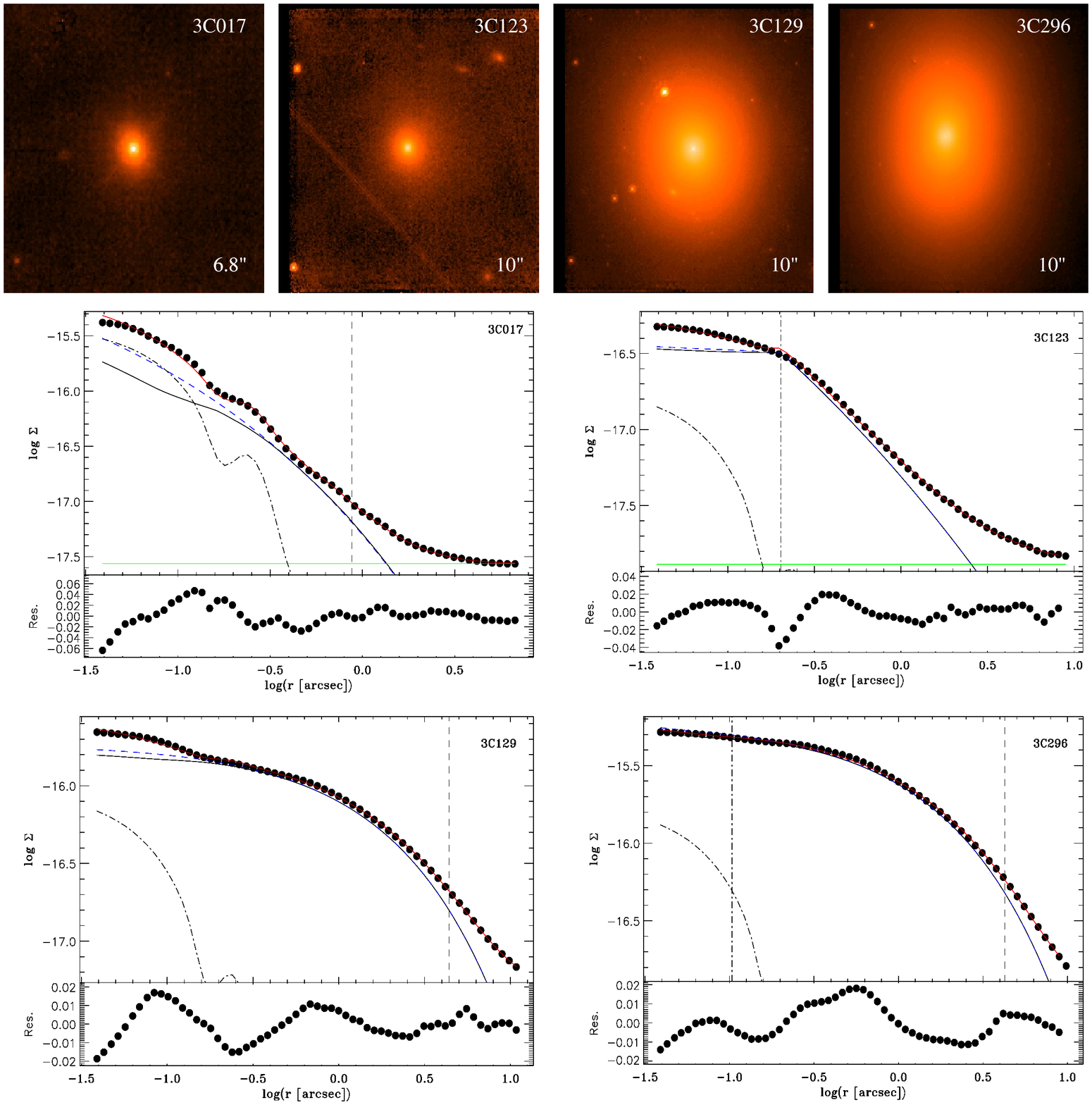} 
\caption{This figure shows four explicative fitting methods used. In the upper
parts we report the NICMOS images of the 4 radio-galaxies. In the lower part
each panel represents the modeling of the surface brightness profile with a
S\'{e}rsic or a core-S\'{e}rsic. The description of lines are as in
Fig.~\ref{esempio}. We mark the position of the effective radius with a black
vertical dot-dashed line and, for a core-S\'{e}rsic model, the break radius
with a dashed vertical line.  More in details we report four examples: in the
upper left panel 3C~17 was fitted with a S\'{e}rsic model; in the upper right
panel 3C~123 with a core-S\'{e}rsic model to obtain an upper-limit to its
nuclear luminosity; in the lower left panel 3C~129 with a S\'{e}rsic model in
the case of an weak nuclear source; and in the lower right panel 3C~296 with a
core-S\'{e}rsic with a weak nuclear point source. In this last panel the
nuclear source plotted in black dashed line is amplified of a factor 10 for a
better visualization. In the plots of 3C~129 and 3C~296 the green horizontal
line of the background level is absent because it is below the range of
intensity shown.}
\label{all}
\end{figure*}

\subsection{Profile fitting}
\label{fitting}

The surface brightness profile can be fitted using a variety of
functions. We used two different models, S\'{e}rsic and core-S\'{e}rsic. The
former \citep{sersic68} is a law of the form:

$$
I(r) = I_{e}  \,\, e^{-b_{n}[(\frac{r}{r_{e}})^{\beta}-1]}
$$
 
where $\beta$ is the concentration parameter and $\beta$ = 1/n, the inverse of
the S\'{e}rsic index n. I$_{e}$ is the intensity at the effective radius
r$_e$.  For a S\'{e}rsic model with 0.5 $\lesssim$ n $\lesssim$ 10, b$_{n}$
$\approx$ 1.9992n - 0.3271 (e.g., \citealt{capaccioli89,caon94,graham05}).

The latter is composed of an outer S\'{e}rsic model and
a inner power-law profile, suitable to describe the innermost
part of the profile when the object shows a deficit
toward the center with respect to the S\'{e}rsic law. \citet{trujillo04}
argued that it is convenient to restrict the core-S\'{e}rsic model to an
infinitely sharp transition between the S\'{e}rsic and power-law profiles
at the break radius r$_{b}$, assuming this form:

$$
I(r) = I(b)\left[\left(\frac{r_{b}}{r}\right)^{\gamma}\theta(r_{b}-r)+e^{b_{n}\left(\left(\frac{r_{b}}{r_{e}}\right)^{\frac{1}{n}}-\left(\frac{r}{r_{e}}\right)^\frac{1}{n}\right)^\frac{1}{n}}\theta(r-r_{b})\right]
$$

where r$_{b}$ separates the inner power law of logarithmic slope $\gamma$ from
the outer part of the profile described by a S\'{e}rsic
model. $\theta$(r$_{b}$-r) is the Heaviside step function.

We fit the brightness profiles with both models by minimizing the residuals,
leaving free to vary the 5 (or 7) variables: the background level, the three
parameters describing the S\'{e}rsic (or five for core-S\'{e}rsic) model and
the nuclear component. Fig.~\ref{all} illustrates four examples of fits to the
NIR brightness profiles. In Sect.~\ref{serocore} we describe the method used
to select the appropriate law for each galaxy.

To take into account the PSF effects, we convolved the models prior to the
comparison with the data \citet{capetti05}. The PSF of NICMOS-NIC2 at the
center of the chip was derived using the TINYTIM program written by
J. Krist. The PSF image was then convolved with a model galaxy with a power
law profile, which represents the innermost region of a
S\'{e}rsic/core-S\'{e}rsic model.  This was done for a set of galaxy models
with different profile slopes $\gamma^{\prime}$, all of them having zero
ellipticity.  We then derived the radial profile of the convolved model
galaxies using circular isophotes, and we calculated the intensity ratio
between the convolved model galaxy and the original model galaxy for different
radii.  With this method we obtained a set of radial ratios for different
values of the innermost slope that can be applied to the models to
``simulate'' the two-dimensional convolution.

The appropriate value of $\gamma^{\prime}$  was measured as part of the
fitting procedure. In the case of a S\'{e}rsic model, $\gamma^{\prime}$ is
the value of the logarithmic slope of the profile at 0$\farcs$1, which
approximately  corresponds to  the resolution  of NICMOS-NIC2.  In the
core-S\'{e}rsic model,  $\gamma^{\prime}$ is the logarithmic  slope of the
inner power-law profile.

\citet{donzelli07} already performed the fitting of the radial brightness
profiles of 3CR sample with S\'{e}rsic law, focusing on their large scale
behavior, thus not including the central region of the galaxies.  For the
objects in which \citet{donzelli07} found a disk-like component in the galaxy
profile, we exclude from the fit the regions dominated by the disk.  The
parameter values found are mostly consistent with those of \citet{donzelli07}
for the objects in which the nuclear component is not dominant with respect to
the galaxy stellar emission. The resulting models produce residuals in the
form of large scale fluctuations with a typical amplitude of $<$ 5\%,
indicative of the goodness of the fit.

Besides the parameters describing the host galaxy profile, the fitting
procedure also yields the flux of the nuclear source.

To assess the reliability of the detection of a nuclear source we adopt the
following operative approach. We extract the radial brightness profile using
the IRAF ``radprof'' task and measure the nuclear FWHM, setting the background
level at the intensity measured just beyond the nuclear point source at a
radius of $\sim$0.3\arcsec. If the FWHM is consistent with that of the PSF
($\sim$ 0.14 arcsec), we took this as an indication of the presence of an
unresolved source.  Alternatively, if there was no evidence for a nuclear
point source, we proceed to fit the profile with a core-S\'{e}rsic model
leaving all parameters free to vary, but fixing $\gamma = 0$. This corresponds
to the most conservative choice for $\gamma$ as it minimizes the galaxy
contribution and maximizes the nuclear source. We consider the nuclear flux
thus derived as the upper limit to the intensity of the central source.

\subsection{S\'{e}rsic or core-S\'{e}rsic?}
\label{serocore}

The choice between S\'{e}rsic and  core-S\'{e}rsic model for a given galaxy is
related  to the corresponding  minimized $\chi^{2}$  value obtained  after the
fitting  procedure.  Since  the   core-S\'{e}rsic  model  includes  more  free
parameters, the  best core-S\'{e}rsic model usually will  reproduce better the
galaxy  profile  than  the  S\'{e}rsic  model.  Therefore,  to  determine  the
goodness  of the  fit, we  use  the ``reduced''  $\chi^{2}$ ($\chi_{r}^{2})  =
\chi^{2}/(n-p) $, where $n$ is the number of fitted data points and $p$ is the
number of  fitted parameters,  which can be  either 3  or 5 for  S\'{e}rsic or
core-S\'{e}rsic  models,  respectively).   We  consider the  best  model  that
corresponds  to the $\chi_{r}^{2}$  closest to  the unity.  Qualitatively, the
choice can be justified by the form of the residuals.  In fact, when fitting a
core-S\'{e}rsic  galaxy with  a pure  S\'{e}rsic model,  the residuals  in the
central  region   show  the  characteristic  S-shape,   already  discussed  by
\citet{trujillo04}.

Nevertheless, since our objects are at distances of up to z = 0.3, we can
misclassify a genuine core galaxy as a pure S\'{e}rsic due to an insufficient
physical resolution of the image. The typical core size of an elliptical
galaxy is $\sim$200 pc \citep[e.g.][]{capetti05,faber97,ferrarese06} which
corresponds to 0$\farcs$14 at z = 0.15. Considering the objects at z $>$ 0.15
with detected nuclei (excluding the BLOs, for which the nucleus outshines the
host galaxy emission in the central region), for which there is ambiguity on
the presence of a core, the typical discrepancy between the estimates of the
nuclear luminosity obtained adopting either a S\'{e}rsic law or a
core-S\'{e}rsic model is $\sim 30$ \%, and at most a factor of 2. The sources
potentially involved in this ambiguity are only $\sim$15\% of the sample and
thus they do not alter the main results of this paper.

\begin{figure}
\includegraphics[scale=0.45]{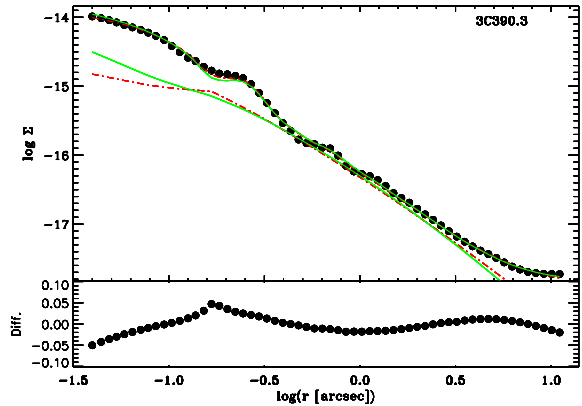}
\caption{The surface brightness profile of 3C~390.3 fitted with a S\'{e}rsic
model (green solid trace) and a core-S\'{e}rsic model (red dot-dashed
trace). The lines below represent the two models used, which include the
PSF-convolved galaxy model, the nuclear source and the background (the green
solid and the red dot-dashed lines correspond to the S\'{e}rsic and
core-S\'{e}rsic, respectively). In the lower panel we show the difference
between the S\'{e}rsic-law and core-S\'{e}rsic models.}
\label{sersiccores}
\end{figure}

The presence of a bright point source may also limit our ability to see a
possible shallow core and to discriminate between S\'{e}rsic and
core-S\'{e}rsic models. However, in those cases (e.g. in the case of the BLOs,
Fig.\ref{sersiccores}), the difference between the nuclear fluxes obtained
adopting the two different models is irrelevant, being smaller than $\sim2$\%,
thus well below the typical photometric error.

\section{Results}

\subsection{Detection of nuclei}
In Table~\ref{table1} we report the 1.6 $\mu$m NIR nuclear
fluxes and luminosities for the sample measured with the procedure explained
in Sect.~\ref{method}. We also give the optical (R band), radio
(nuclear at 5 GHz and large scale at 178 MHz) fluxes, and [O III] emission
line luminosities for all the objects of the sample.

\begin{table*}
  \begin{center}
    \caption{Basic data for all 3CR source considered}
    \label{table1}
\begin{tabular}{l|ccc|ccc|ccc|cc|c}
\tablewidth{0pt}
\hline
\hline
Name & z & FR - E.L. & Ref & log F$_{178 MHz}$ & log F$_{r}$ & Ref & log F$_{o}$ & log F$_{IR}$ & log L$_{IR}$ & log L$_{[O~III]}$ & Ref & Profile \\
\hline
   3C~15&    0.073 &     LIG     &B09 & -21.80 &  -23.42  &Z95 & $<$-28.05 & -27.42    &  27.55    &  40.46   &B09& S\\
   3C~17&   0.2197 &     BLO     &B09 &-21.70  &  -23.08  &Z95 & -26.91    & -26.84    &  29.00    &  41.68   &B09& S\\  
   3C~20&    0.174 &     HIG     &J97 &-21.37  &  -25.48  &F97 &   -       & $<$-27.72 &  $<$27.94 & 41.47    &L96& cS\\
   3C~28&   0.1953 &    ELEG     &B09 &-21.79  &$<$-26.70 &G88 & $<$-29.28 & $<$-29.64 &  $<$26.10 & 40.67    &B09& cS\\
   3C~29&   0.0488 &    FR~I     &B09 &-21.82  &  -24.04  &M93 & -28.02    & -27.56    &  27.03    & 40.06    &B09& S\\
   3C~31&    0.017 &    FR~I     &B09 &-21.77  &  -24.04  &G88 & -27.61    & -27.44    &  26.29    & 39.41    &B09& cS\\
   3C~33&    0.059 &     HIG     &B09 &-21.28  &  -24.52  &Z95 &   -       & $<$-27.14 &  $<$27.64 & 42.05    &B09& cS\\
 3C~33.1&   0.1809 &     BLO     &B09 &-21.89  &  -24.66  &Z95 &  -27.5    & -27.12    &  28.57    & 42.03    &B09& S\\
   3C~35&    0.067 &     LIG     &B09 &-21.98  &  -24.62  &Z95 & $<$-28.41 & $<$-28.30 &  $<$26.59 & 39.87    &B09& cS\\
   3C~52&  0.28540 &     LIG     &B09 &-21.87  &  -25.11  &C08 &    -      & $<$-28.85 &  $<$27.18 & $<$40.84 &B09& cS\\
 3C~61.1&     0.184&     HIG     &B09 &-21.51  &  -25.43  &Z95 &   -       & $<$-28.57 &  $<$27.13 & 42.20    &B09& cS\\
  3C~66B&   0.0215 &    FR~I     &B09 &-21.61  &  -23.74  &G88 & -27.09    & -27.04    &  26.91    & 39.97    &B09& cS\\
  3C~75N&  0.02215 &    FR~I     &B09 &-21.59  &  -24.29  &Z95 &   -       & $<$-28.48 &  $<$25.48 & $<$39.80 &B09& cS\\
 3C~76.1&  0.03249 &    FR~I     &B09 &-21.88  &  -24.61  &G88 &   -       & $<$-28.29 &  $<$25.99 & $<$39.75 &B09& cS\\
   3C~79&    0.256 &    HIG      &B09 &-21.52  &  -24.76  &Z95 & -27.62    & -27.44    &  28.51    &  42.51   &B09& S\\
 3C~83.1&   0.0251 &    FR~I     &B09 &-21.58  &  -24.67  &R75 & -28.64    & -27.65    &  26.43    & $<$39.33 &B09& cS\\
   3C~84&  0.01756 &    FR~I     &B09 &-21.39  &  -21.37  &T96 & -25.61    & -25.47    &  28.28    & 41.53    &B09& S\\
   3C~88&  0.03022 &    LIG      &B09 &-21.82  &  -23.72  &Z95 & -27.62    & -27.46    &  26.76    & 40.05    &B09& S\\
   3C~98&     0.03 &    HIG      &B09 &-21.45  &  -24.97  &Z95 & $<$-28.55 & $<$-27.64 &  $<$26.58 & 40.90    &B09& cS\\
  3C~105&    0.089 &    HIG      &B09 &-21.75  &  -24.76  &Z95 &   -       & -27.82    &   27.3    & 41.29    &B09& S\\
  3C~111&   0.0485 &    BLO      &B09 &-21.19  &  -22.93  &Z95 & -26.48    & -25.30    &  29.33    & 42.33    &B09& S\\
  3C~123&   0.2177 &    LIG      &B09 &-20.72  &  -24.01  &H98b& $<$-29.39 & $<$-28.16 &  $<$27.67 & 41.69    &B09& cS\\
  3C~129&   0.0208 &    FR~I     &B09 &-21.33  &  -24.99  &V82 &   -       & -27.48    &  26.43    & $<$39.75 &B09& S\\
3C~129.1&  0.02220 &    FR~I     &B09 &-21.98  &  -25.51  &C08 &   -       & $<$-28.52 & $<$25.44  & $<$40.39 &B09& cS \\
  3C~130&    0.032 &    FR~I     &H98a&-21.81  &  -23.48  &H99 &   -       & $<$-28.04 &  $<$26.24 & $<$40.59 &B09& cS\\
  3C~132&    0.214 &    LIG      &J97 &-21.91  &$<$-24.42 &Z95 & $<$-28.99 & $<$-28.03 &  $<$27.78 &  -       &-  & cS\\
  3C~133&   0.2775 &    HIG      &B09 &-21.65  &  -23.69  &N98 & -28.01    & -26.82    &  29.18    & 42.39    &B09& S\\
  3C~135&   0.1253 &    HIG      &B09 &-21.76  &  -25.22  &Z95 & -28.04    & -27.60    &  27.80    & 41.85    &B09& S\\
  3C~153&    0.277 &    LIG      &B09 &-21.78  &$<$-26.28 &H98b& $<$-29.35 & $<$-27.94 &  $<$28.07 & 41.26    &B09& cS\\
  3C~165&   0.2957 &    LIG      &B09 &-21.87  &  -24.97  &H98b& -28.77    & -28.39    &  27.67    & 41.28    &B09& S\\
  3C~166&    0.245 &    LIG      &B09 &-21.83  &  -23.19  &Z95 & -27.91    & -27.28    &  28.63    & 41.32    &B09& S\\
  3C~171&   0.2384 &    HIG      &B09 &-21.71  &  -25.54  &Z95 & -29.40    & $<$-28.05 &  $<$27.84 & 42.55    &B09& cS\\
3C~173.1&   0.2921 &    LIG      &B09 &-21.81  &  -24.86  &Z95 &   -       & -28.18    &  27.86    & 40.47    &B09& cS\\
3C~180  &   0.2200 &    HIG      &B09 &-21.82  &    -     & -  &   -       & $<$-28.50 & $<$27.34  & 42.03    &B09& cS\\
3C~184.1&    0.1182&    BLO      &B09 &-21.89  &  -25.10  &Z95 & -27.18    & -26.60    &  28.76    & 42.02    &B09& S\\
  3C~192&   0.0598 &    HIG      &B09 &-21.64  &  -25.07  &Z95 & $<$-27.81 & $<$-27.92 &  $<$26.88 & 41.22    &B09& cS\\
3C~196.1&    0.198 &    LIG      &B09 &-21.73  &  -24.11  &B89 &   -       & $<$-28.57 &  $<$27.18 & 41.23    &B09& cS\\
3C~197.1&    0.1301&    BLO      &B09 &-22.09  &  -25.13  &Z95 & -27.33    & -27.32    &  28.11    & 40.73    &B09& S\\
3C~198  &   0.0815 &    SFG      &B09 &-22.01  &$<$-25.77 &F78 & -27.02    & -26.96    &  28.09    & 40.82    &B09& S\\
3C~213.1&   0.1937 &    LIG      &B09 &-22.18  &  -23.22  &A95 &   -       & -27.52    &  28.22    & 40.78    &B09& S\\
  3C~219&   0.1744 &    BLO      &B09 &-21.39  &  -24.12  &Z95 & -27.02    & -26.95    &  28.71    & 41.51    &B09& S\\
  3C~223&   0.1368 &    HIG      &B09 &-21.83  &  -24.90  &Z95 & $<$-28.20 & -27.56    &  27.91    & 41.95    &B09& S\\
3C~223.1&    0.107 &    HIG      &B09 &-22.22  &  -25.03  &H98b& $<$-28.09 & -26.65    &  28.62    & 41.39    &B09& S\\
  3C~227&   0.0861 &    BLO      &B09 &-21.52  &  -24.62  &Z95 & -26.26    & -25.70    &  29.40    & 41.60    &B09& cS\\
  3C~234&    0.185 &    HIG      &B09 &-21.58  &  -23.83  &Z95 & -26.61    & -26.04    &  29.66    & 42.84    &B09& S\\
  3C~236&   0.1005 &    LIG      &B09 &-21.84  &  -23.71  &Z95 & $<$-28.18 & -27.50    &  27.72    & 40.72    &B09& S\\
  3C~258&   0.1650 &    LIG      &B09 &-22.01  &  -       & -  &    -      & $<$-28.38 &  $<$27.24 & 39.94    &B09& cS\\
  3C~264&   0.0206 &    FR~I     &B09 &-21.58  &  -23.70  &G88 & -26.62    & -26.64    &  27.28    & 39.07    &B09& cS\\
  3C~270&   0.0074 &    FR~I     &B09 &-21.26  &  -23.51  &G88 & -27.97    & -27.05    &  25.96    & 37.89    &B09& cS\\
3C~272.1&   0.0037 &    FR~I     &B09 &-21.71  &  -23.75  &G88 & -26.88    & -27.07    &  25.35    & 38.13    &B09& cS\\
  3C~274&   0.0037 &    FR~I     &B09 &-19.98  &  -22.40  &G88 & -26.06    & -25.52    &  26.90    & 38.80    &B09& cS\\
3C~277.3&    0.0857&    HIG      &B09 &-22.05  &  -24.91  &G88 & -28.61    & $<$-27.08 &  $<$28.09 & 40.78    &B09& cS\\
  3C~284&    0.239 &    HIG      &B09 &-22.21  &  -25.65  &Z95 &   -       & $<$-28.53 &  $<$27.37 & 41.26    &B09& cS\\
  3C~285&   0.0794 &    HIG      &B09 &-21.95  &  -25.10  &Z95 & -29.38    & -27.93    &  27.11    & 40.40    &B09& S\\
3C~287.1&   0.2159 &    BLO      &B09 &-22.09  &  -23.29  &Z95 & -27.11    & -26.41    &  29.42    & 41.43    &B09& S\\
  3C~288&    0.246 &    FR~I     &Z95 &-21.72  &  -24.53  &G88 & -28.94    & -28.71    &  27.40    &  -       & - & cS\\
  3C~293&   0.0452 &    LIG      &B09 &-21.90  &  -24.81  &G88 &   -       & -28.09    &  26.47    & 39.70    &B09& cS\\
  3C~296&   0.0237 &    FR~I     &B09 &-21.89  &  -24.10  &G88 & -28.25    & -28.13    &  25.92    & 39.65    &B09& cS\\
  3C~300&     0.27 &    HIG      &B09 &-21.75  &  -24.92  &Z95 & -28.20    & -27.83    &  28.16    & 41.64    &B09& S\\
  3C~303&     0.141&    BLO      &B09 &-21.95  &  -23.69  &Z95 & -26.83    & -26.49    &  29.01    & 41.51    &B09& S\\
 3C~305 &   0.04164&    HIG      &B09 &-21.80  &  -24.50  &G88 &   -       & $<$-28.10 & $<$26.40  & 40.93    &B09& cS\\
  3C~310&    0.054 &    FR~I     &B09 &-21.26  &  -24.10  &G88 & -28.24    & $<$-27.07 &  $<$27.69 & 39.95    &B09& cS\\
3C~314.1&    0.1197&    ELEG     &B09 &-21.98  &$<$-26.00 &G88 & $<$-28.81 & $<$-27.37 &  $<$28.09 & 39.49    &B09& cS\\
  3C~315&   0.1083 &    FR~I     &B09 &-21.75  &  -23.75  &G88 &   -       & -27.76    &  27.52    & 40.69    &B09& cS\\
  3C~317&   0.0342 &    FR~I     &B09 &-21.33  &  -23.41  &M93 & -27.67    & -27.05    &  27.31    & 40.25    &B09& cS\\
  3C~319&    0.192 &    LIG      &B09 &-21.82  &$<$-25.80 &Z95 & $<$-28.33 & $<$-27.77 &  $<$27.96 & $<$39.91 &B09& cS\\
  3C~321&    0.096 &    HIG      &B09 &-21.87  &  -24.41  &Z95 &   -       & -27.64    &  27.55    & 40.73    &B09& cS\\
3C~323.1&    0.264 &    BLO      &B09 &-22.00  &  -24.28  &Z95 & -25.83    & -25.77    &  30.20    & 42.45    &B09& S\\
  3C~326&     0.089&    LIG      &B09 &-22.11  &  -24.78  &Z95 & $<$-28.04 & -27.87    &  27.25    & 40.23    &B09& cS\\
  3C~332&   0.1515 &    BLO      &B09 &-22.02  &  -24.90  &Z95 & -26.76    & -26.08    &  29.47    & 41.57    &B09& S\\
  3C~338&  0.03035 &    FR~I     &B09 &-21.33  &  -23.99  &G88 & -27.78    & -28.17    &  26.08    & 39.44    &B09& cS\\
\hline
  \multicolumn{12}{c}{{Continued on Next Page}} \\
    \end{tabular}
  \end{center}
\end{table*}

\addtocounter{table}{-1}
\begin{table*}
  \begin{center}
    \caption{Continued}
\begin{tabular}{l|ccc|ccc|ccc|cc|c}
\tablewidth{0pt}
\hline
\hline
Name & z & FR - E.L. & Ref & log F$_{178 MHz}$ & log F$_{r}$ & Ref & log F$_{o}$ & log F$_{IR}$ & log L$_{IR}$ & log L$_{[O~III]}$ & Ref & Profile \\
\hline
  3C~346&    0.162 &   FR~I      &S91 &-21.96  &  -23.66  &G88 & -27.42    & -27.01    &  28.73    & 41.59    &G94& S\\
  3C~348&    0.154 &   ELEG      &B09 &-20.45  &  -25.00  &M93 & -28.88    & $<$-27.34 &  $<$28.34 & 40.17    &B09& cS\\
  3C~349&    0.205 &   HIG       &J97 &-21.88  &  -24.60  &H98b& -27.62    & -27.39    &  28.39    & 41.82    &R89& S\\
  3C~353&  0.03043 &    LIG      &B09 &-20.63  &  -23.69  &Z95 & $<$-28.70 & $<$-27.99 &  $<$26.24 & 40.05    &B09& cS\\
  3C~357&    0.167 &    LIG      &B09 &-22.09  &  -25.15  &Z95 & $<$-28.73 & $<$-27.94 &  $<$27.68 & 40.70    &B09& cS\\
  3C~371&    0.051 & BLLac-LIG   &B09 &-22.43  &  -23.04  &P81 & -25.56    & -24.91    &  29.75    & 40.84    &B09& S\\
3C~379.1&    0.256 &    HIG      &B09 &-22.13  &  -25.26  &S85 & $<$-28.74 & $<$-28.05 &  $<$27.89 & 41.51    &B09& cS\\
  3C~381&   0.1605 &    HIG      &B09 &-21.78  &  -25.11  &Z95 & $<$-28.08 & -27.24    &  28.36    & 42.12    &B09& S\\
  3C~382&   0.0578 &    BLO      &B09 &-21.70  &  -23.64  &Z95 & -25.05    & -24.91    &  29.86    & 41.65    &B09& S\\
  3C~388&    0.091 &    LIG      &B09 &-21.61  &  -24.10  &Z95 & -27.88    & -27.83    &  27.31    & 40.54    &B09& S\\
3C~390.3&   0.0561 &    BLO      &B09 &-21.32  &  -23.36  &Z95 & -25.70    & -25.42    &  29.32    & 41.96    &B09& S\\
  3C~401&    0.201 &    LIG      &B09 &-21.68  &  -24.27  &Z95 & -28.15    & -28.26    &  27.51    & 40.76    &B09& S\\
  3C~402&   0.0239 &    LIG      &B09 &-22.00  &  -24.29  &Z95 & -27.43    & $<$-27.87 &  $<$26.15 & $<$39.34 &B09& cS\\
  3C~403&    0.059 &    HIG      &B09 &-21.75  &  -24.92  &Z95 & -28.14    & -26.97    &  27.81    & 41.62    &B09& S\\
3C~403.1&   0.05540 &   LIG      &J97 &-21.87  &   -      & -  &     -     & $<$-27.97 & $<$26.76  & -        & - & cS\\
  3C~405&  0.05607 &    HIG      &O75 &-19.06  &  -23.46  &Z95 &   -       & -27.84    &  26.90    & 41.96    &O75& cS\\
  3C~410&  0.248500 &   FR~II    &D96 & -21.46 &  -22.83  & C08&   -       & -26.25    &  29.68    &  -       & - & S\\
  3C~424&  0.126988&    FR~I     &B09 &-21.85  &  -24.67  &B92 &$<$-28.61  & -27.77    &  27.64    & 40.60    &B09& S\\
  3C~430&   0.0556 &    LIG      &B09 &-21.47  &  -24.77  &S84 &   -       & $<$-27.60 &  $<$27.14 & 40.23    &B09& cS\\
  3C~433&   0.1016 &    HIG      &B09 &-21.25  &  -25.24  &G88 &   -       & -26.51    &  28.73    & 41.50    &B09& S\\
  3C~436&   0.2145 &    HIG      &B09 &-21.75  &  -24.61  &Z95 &   -       & $<$-28.06 &  $<$27.76 & 41.25    &B09& cS\\
  3C~438&     0.29 &    LIG      &B09 &-21.35  &  -24.60  &Z95 & $<$-29.18 & $<$-28.28 &  $<$27.76 & $<$41.08 &B09& cS\\
  3C~442&   0.0263 &    FR~I     &B09 &-21.79  &  -25.68  &G88 & -28.82    & -27.75    &  26.35    & 39.14    &B09& cS\\
  3C~445&    0.057 &    BLO      &B09 &-21.61  &  -23.42  &Z95 & -25.49    & -25.31    &  29.45    & 42.39    &B09& S\\
  3C~449&  0.01708 &    FR~I     &B09 &-21.94  &  -24.44  &G88 & -27.53    & -27.22    &  26.58    & 39.13    &B09& S\\
  3C~452&   0.0811 &    HIG      &B09 &-21.26  &  -23.81  &Z95 & -28.16    & $<$-27.89 &  $<$27.16 & 41.19    &B09& cS\\
  3C~459&   0.2199 &    BLO      &B09 &-21.59  &  -22.82  &Z95 &   -       & -26.65    &  29.18    & 41.73    &B09& S\\
  3C~465&   0.0301 &    FR~I     &B09 &-21.42  &  -23.57  &G88 & -27.50    & -27.39    &  26.86    & 39.71    &B09& cS\\
\hline					   		   			 		         	   
NGC~6251&  0.02471 &    FR~I     &P84 &-22.00  &  -23.05  &J86 & -26.82    & -26.28    &  27.80    & 39.70    &S81& cS\\
\hline
    \end{tabular}
  \end{center}
Description of the table: Col.(1): name; Col.(2): redshift; Col.(3): Fanaroff
\& Riley and optical spectral classifications. For simplicity, we mark as FR~I
all radio galaxies with FR~I-like radio morphology and LIG-like optical
spectrum, as LIG all radio galaxies with FR~II-like radio morphology and
LIG-like optical spectrum, as HIG (BLO) all radio galaxies with FR~II-like
radio morphology and HIG-like optical spectrum with the absence (presence) of
broad emission lines. References (Col. 4): B09 \citet{buttiglione10}, J97
\citet{jackson97}, H98a \citet{hardcastle98a}, Z95 from \citet{zirbel95}, S91
\citet{spencer91}, O75 \citet{osterbrock75}, D96 \citet{dekoff96}, P84
\citet{perley84}; Col.(5): flux density at 178 MHz in erg s$^{-1}$ cm$^{-2}$
Hz$^{-1}$ (in these logarithmic units -23 corresponds to 1 Jy); Col.(6):
nuclear radio flux density at 5 GHz in erg s$^{-1}$ cm$^{-2}$
Hz$^{-1}$. References (Col. 7): B92 \citet{black92} (8.3 GHz), G88:
\citet{giovannini88}, L91: \citet{leahy91}, M93: \citet{morganti93}, R75
\citet{riley75}, T96 \citet{taylor96}, Z95 \citet{zirbel95}, N98
\citet{nilsson98}, H98b \citet{hardcastle98b} (8.4 GHz) F97:
\citet{fernini97}, V82: \citet{vanbreugel82}, H99: \citet{hardcastle99}, B89:
\citet{baum89}, F78: \citet{fomalont78}, A95: \citet{akujor95}, P81:
\citet{pearson81}, S85: \citet{spangler85} (1.5GHz), S84: \citet{spangler84},
J86: \citet{jones86}, C08: A. Capetti, priv. com.. Radio core data at
frequencies different from 5 GHz were converted to 5 GHz using a flat
($\alpha$ = 0) spectral index; Col.(8): nuclear optical flux density in erg
s$^{-1}$ cm$^{-2}$ Hz$^{-1}$ (\citealt{chiaberge99}, \citealt{chiaberge00} and
\citealt{chiaberge02}) in R band; Col.(9): nuclear 1.6 $\mu$m near infrared
flux density in erg s$^{-1}$ cm$^{-2}$ Hz$^{-1}$; Col.(10): nuclear 1.6 $\mu$m
near infrared luminosity in erg s$^{-1}$ Hz$^{-1}$; Col.(11):
[O~III]$\lambda\lambda$5007 luminosity in erg s$^{-1}$. References (Col.12):
B09 \citet{buttiglione09}, O75 \citet{osterbrock75}, S81 \citet{shuder81}, L96
\citet{lawrence96}, R89 \citet{rawlings89}. Col.(13): type of model used to
fit the NIR surface brightness profile.
\end{table*}

Here we examine the detection rate of the NIR central sources
(Table~\ref{detect}),  considering each radio-morphological and
optical-spectral class separately. Overall, we detect nuclei in 64\%$\pm$5\%
of the entire 3CR sample. We detect nuclei in 81\%$\pm$8\% and 58\%$\pm$6\% of
the FR~Is and FR~IIs, respectively. Among the FR~IIs, all of the BLOs have an
unresolved nuclear point source, while the detection rate of nuclei in LIGs
and HIGs is 44\%$\pm$10\% and 54\%$\pm$9\%, respectively. The only one BL~Lac
included in the sample has a bright nuclear source. The nucleus of 3C~410, the
only spectroscopically unclassified FR~II, is detected, while neither the
ELEGs nor the star forming galaxy 3C~198 have an NIR nucleus.  The different
detection rates in the FR~Is and FR~IIs are not a consequence of the redshift
distribution of the 3CR catalog. All FR~II sources (HIG, LIG, and BLO) in our
sample share a similar median redshift, and only the FR~Is have a lower median
$z$, because of their lower total power.  

Due to the presence of upper limits, we use the Kaplan-Meier
product-limit estimator \citep{kaplan58}, which provides the mean value for a
distribution with censored data, where the censoring is random, to measure the
average value of NIR nuclear luminosities of each class (reported in
Table~\ref{detect}). The BLOs have, on average, the brightest near
infrared nuclear sources in the sample in luminosity ($<$L$_{IR}$$>$ =
1.6$\times$10$^{29}$ erg s$^{-1}$ Hz$^{-1}$). The FR~I nuclei are on average
the faintest ($<$L$_{IR}$$>$ = 4.0$\times$10$^{26}$ erg s$^{-1}$ Hz$^{-1}$).
FR~II LIGs and HIGs share a similar range of NIR luminosity and appear to
bridge the gap between BLOs and FR~Is.  Undetected NIR nuclei are
present for F$_{IR}<$ 10$^{-27}$ erg cm$^{-2}$ s$^{-1}$ Hz$^{-1}$ and
L$_{IR}<$ 2.5$\times$10$^{28}$ erg s$^{-1}$ Hz$^{-1}$.  However about one half
of the detected nuclei have fluxes and luminosities below these values.

\begin{table}
\begin{center}
\caption{Detections of nuclei in the sample}
\begin{tabular}{c|cccc |c}
\tableline\tableline
Class & Nuclei & Upper lim. & Tot. & $\%$ & Log L$_{IR}$ \\
\tableline
FR~I  &  21  & 5  & 26 & 81 & 26.7 $\pm$  0.2 \\
\tableline
FR~II &  42  & 31 & 73 & 58 & $-$ \\
\tableline
LIG   & 11   & 14 & 25 & 44  & 26.9 $\pm$  0.2\\
HIG   & 15   & 13 & 28 & 54  & 27.5 $\pm$  0.2\\
BLO   & 15   & 0  & 15 & 100 & 29.2 $\pm$  0.1\\
\tableline
ELEG  &  0   & 3  & 3  & 0   & $-$ \\
SFG   &  0   & 1  & 1  & 0   & $-$ \\
UNDEF & 1    & 0  & 1  & 100 & $-$ \\
\tableline
BLLac  & 1   & 0  & 1  & 100 & $-$ \\
\tableline	       	   
\tableline
Tot.  & 64   & 36 & 100 & 64  & \\
\tableline
\end{tabular}
\label{detect}
\tablecomments{Detection rate of 1.6 $\mu$m near infrared nuclei for
  each class. The ELEG (Extremely Low Ionization Galaxies) are 3C~28,
  3C~314.1, and 3C~348. The BL Lac is 3C~371, the Star forming Galaxy (SFG) is
  3C~198 and the Undefined object (UNDEF) is 3C~410 because of the lack of
  spectral data. LIGs, HIGs, BLOs, ELEGs, SFG and UNDEF belong to FR~II class.
  Column description: (1) class; (2) number of detected nuclei; (3) number of
  undetected nuclei; (4) total number of objects for each class; (5)
  percentage of detection; (6) the mean value and their errors of the NIR
  luminosities for each class with the Kaplan-Meier statistics. }
\end{center}
\end{table}

\subsection{Nuclei in the optical-infrared planes}

\begin{figure*}
\includegraphics[scale=0.30]{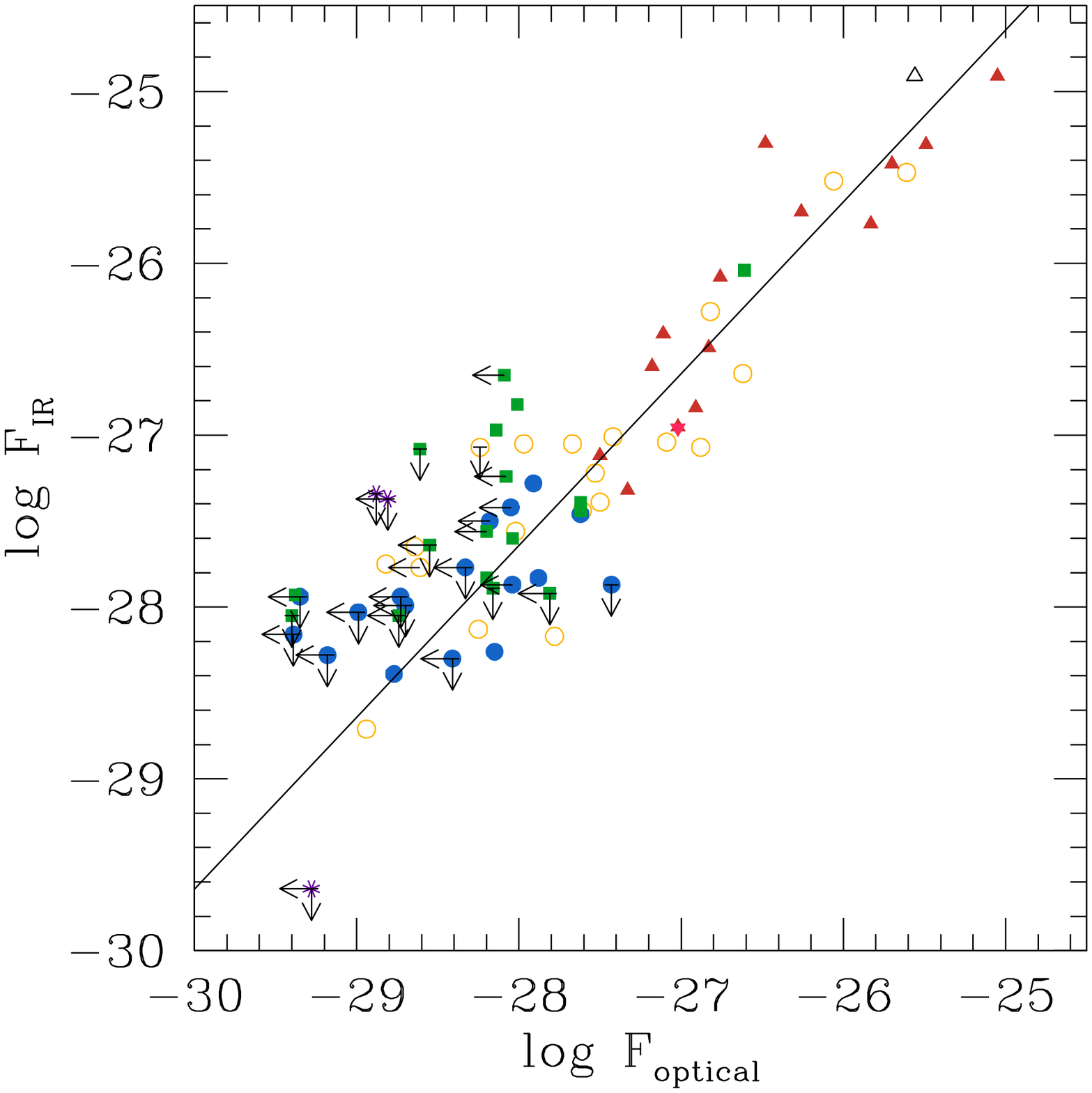}
\includegraphics[scale=0.30]{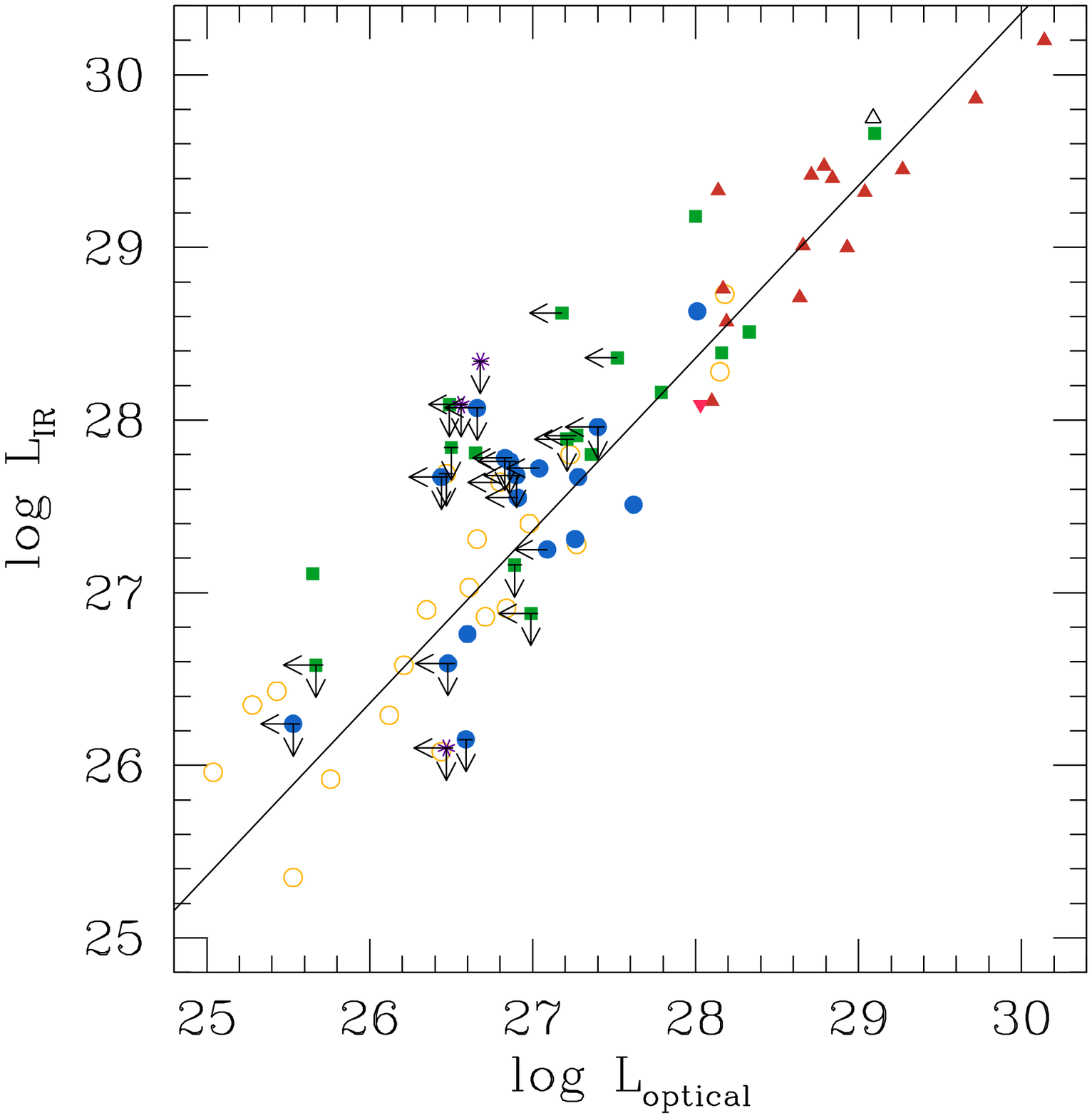}
\includegraphics[scale=0.30]{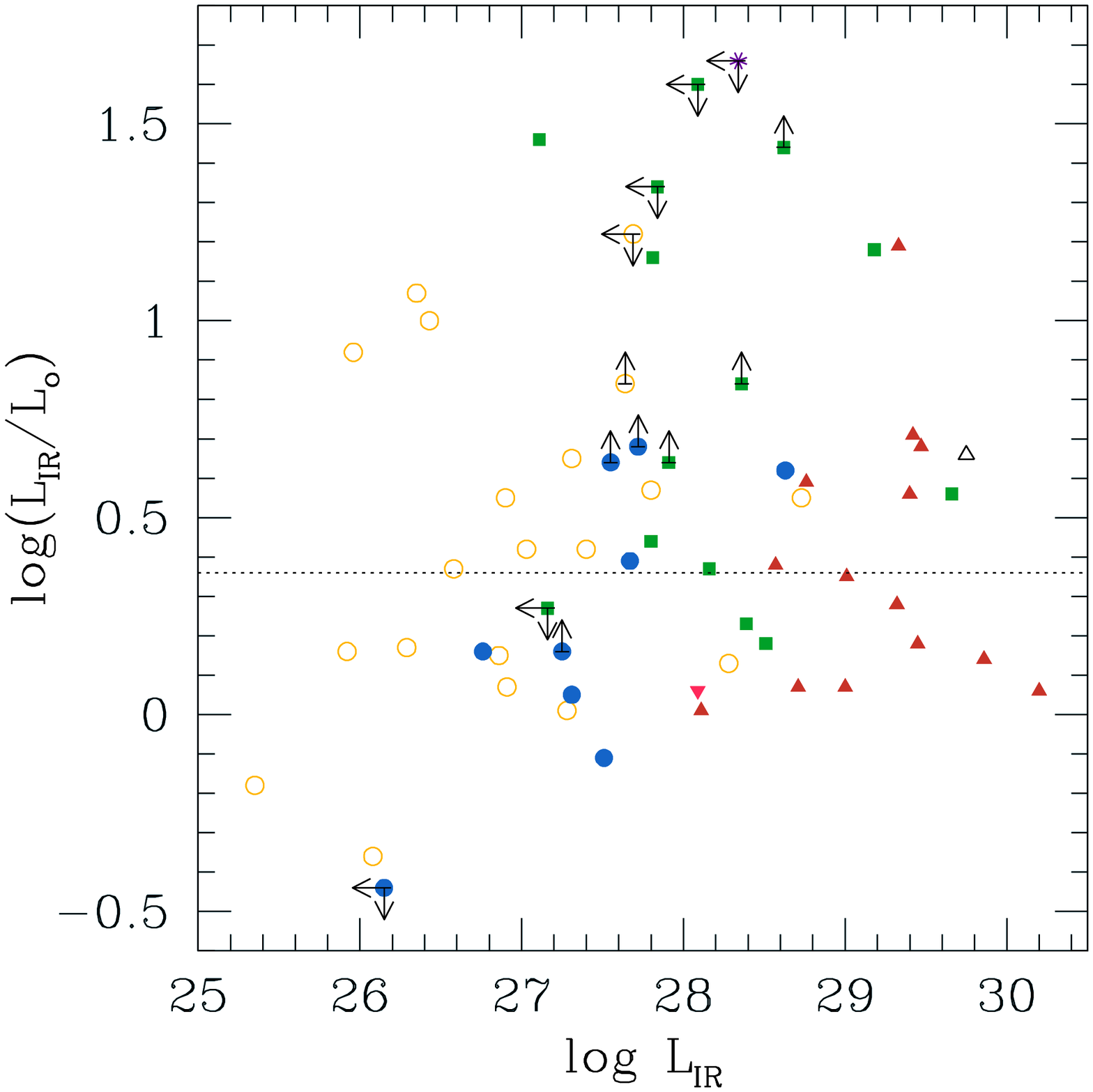}
\caption{In left panel, log$F_{o}$-log$F_{IR}$ (in units of erg s$^{-1}$
  cm$^{-2}$ Hz$^{-1}$); in median panel, log$L_{o}$-log$L_{IR}$; in right
  panel, log$L_{IR}$/$L_{o}$-log$L_{IR}$ (in units of erg s$^{-1}$
  Hz$^{-1}$). In these plots the empty yellow circles represent the FR~Is, the
  filled blue circles the FR~II LIGs, the filled green squares the HIGs and
  the filled red triangles the BLOs, the purple asterisks the ELEGs, the empty
  black triangle the BL Lac (3C~371), and the filled pink reversed triangle
  the SFG (3C~198). The solid or dashed lines present in these plots are the
  line representing the unity of the ratio $\nu$L$_{IR}$/$\nu$L$_{o}$. In
  the right-panel plot, although the same factor L$_{IR}$ appears in both
  axes, its aim is not to find a relation or a trend, but to better point out
  the outliers in log($L_{IR}$/$L_{o}$).}
\label{oir}
\end{figure*}

It is instructive to compare the NIR properties with those of the optical
nuclei.  In Fig.~\ref{oir}, left panel, we plot F$_{\rm o}$-F$_{IR}$ and
L$_{\rm o}$-L$_{IR}$ for all objects in the sample.  Not surprisingly, because
of the relative proximity of the two bands, most of the points occupy a narrow
strip parallel to the bisectrix of the planes. However, we have to be
cautious in taking this plot at face value as it is sensitive to the presence
of nuclear absorption and also to nuclear variability, since infrared and
optical data were not obtained simultaneously.

The FR~I galaxies (empty yellow circles) cover the whole range in optical and
NIR fluxes ($\sim$5 dex). The FR~II LIGs (filled blue circles) are clustered
in small optical and NIR flux range ($\sim$2 dex) in the lower-left slide of
the plane. The HIGs (filled green squares ) are also clustered, but with a
slight tail towards higher flux, covering a wider range in both bands ($\sim$3
dex) with respect to FR~II LIGs. The BLOs (filled red triangles) show the
highest optical and NIR fluxes of the sample.

In the L$_{O}$-L$_{IR}$ plane (Fig.~\ref{oir}, middle panel) we find that FRIs
exhibit the lowest luminosities. FR~II LIGs and HIGs are present at
increasing optical and NIR
luminosities. Finally,  the BLOs are by far the most powerful objects.

To highlight the presence of outliers in the L$_{O}$-L$_{IR}$ plane, in
Fig.~\ref{oir} (right panel) we show the ratio between NIR and optical nuclear
luminosities plotted vs.  the NIR nuclear luminosity.  Most objects lie within
$\sim$0.5 dex of the dashed line representing $\nu$L$_{IR}$/$\nu$L$_{o} =
1$. Note that most of the nuclei that are found above this line (i.e. showing
an NIR excess with respect to the optical emission) belong to the HIG class.
We discuss the origin of this behavior of the HIGs in more detail in
Sect.~\ref{discusHIG}.

\subsection{Nuclei in the radio-infrared planes}

\begin{figure*}
\includegraphics[scale=0.45]{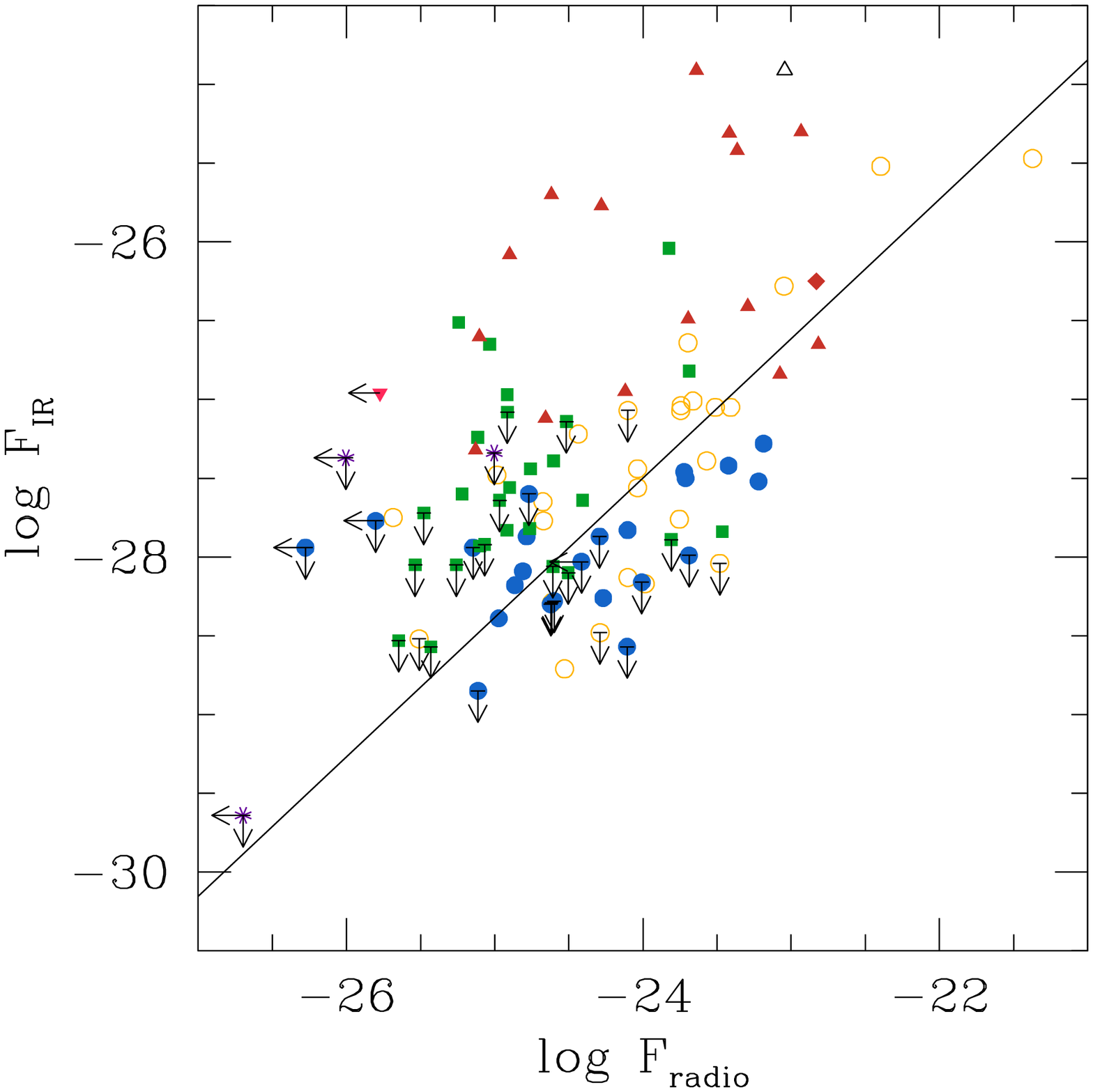}
\includegraphics[scale=0.45]{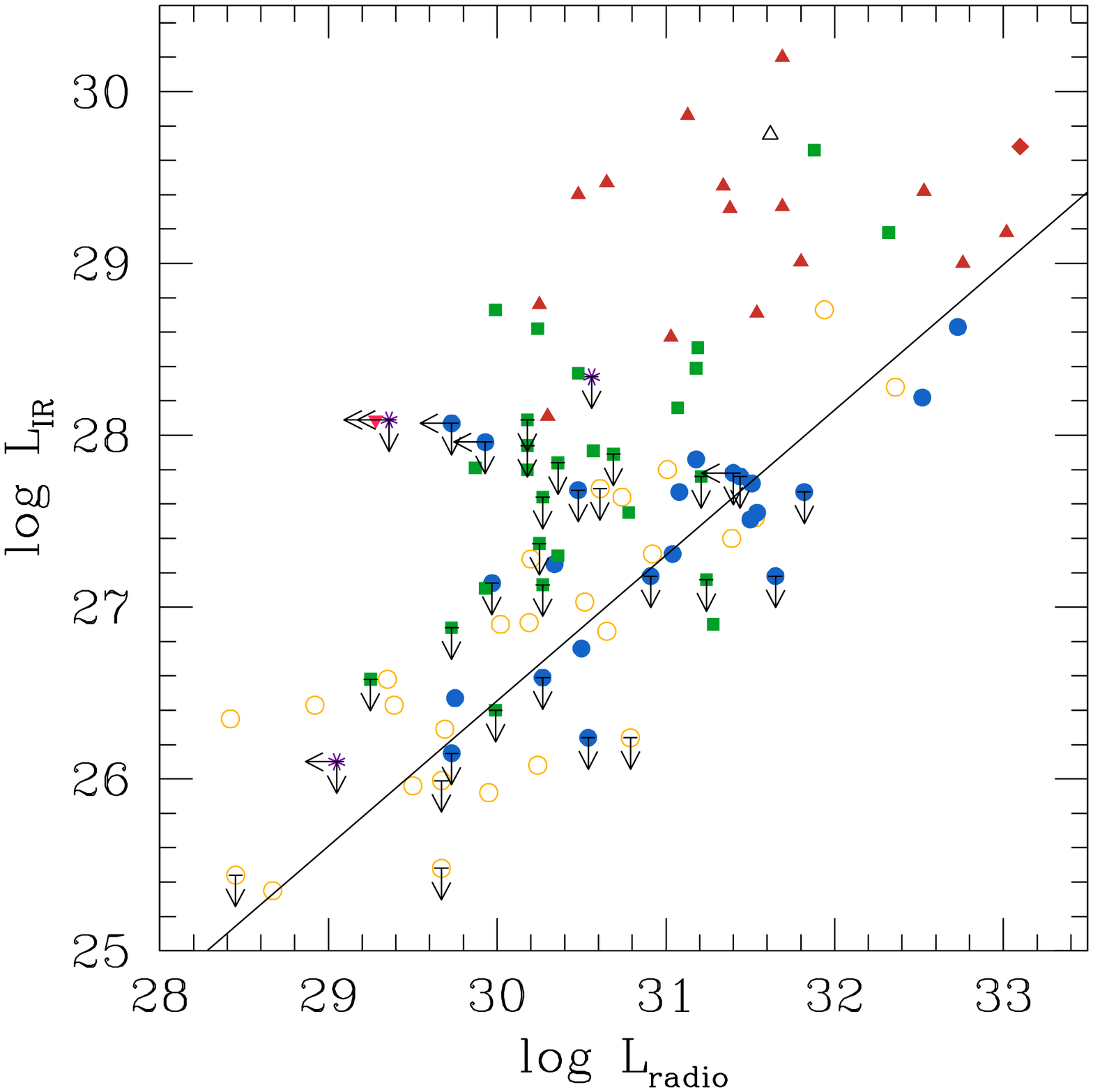}
\caption{Left panel: radio core flux vs. near infrared nuclear flux (in units
of erg s$^{-1}$ cm$^{-2}$ Hz$^{-1}$). Right panel: radio core luminosity
vs. near infrared nuclear luminosity (in units of erg s$^{-1}$ Hz$^{-1}$). In
these plots the empty yellow circles represent the FR~Is, the filled blue
circles the FR~II LIGs, the filled green squares the HIGs and the filled red
triangles the BLOs, the purple asterisks the ELEGs, the empty black triangle
the BL Lac (3C~371), the filled pink reversed triangle the SFG (3C~198), and
the filled red diamond is the undefined object (3C~410). The solid line
represents the linear correlation found for FR~Is and FR~II LIGs.}
\label{rir}
\end{figure*}

In order to investigate the origin of the NIR nuclei, we now explore possible
correlations between the properties of the infrared nuclear emission and those
of the radio cores.  Fig.~\ref{rir}, left panel, shows the IR nuclear flux
F$_{IR}$ plotted vs. the radio core flux F$_{r}$. The FR~Is (empty yellow
circles) cover the whole range of flux spanned by the sample, both in the
radio and in the NIR. HIGs (filled green squares) and FR~II LIGs (filled blue
circles) are clustered in the central region of the plane. The BLOs (filled
red triangles) are located in the top-right quadrant, at higher radio and NIR
fluxes.  A similar distribution of the points is evident in the plane defined
by the radio and the NIR core luminosities (Fig.~\ref{rir}, right
panel). However in the L$_{r}$-L$_{IR}$ plane the distributions are stretched
because of the common dependence of the two quantities on distance. The FR~I
galaxies have radio core luminosities that are on average lower than those of
FR~IIs.  This is expected because the Fanaroff-Riley morphological radio
classification in FR~I and FR~II corresponds to different total luminosities
at 178 MHz, and a positive trend links total and core radio luminosity
\citep{giovannini88}.

A clear trend is visible for FR~Is both in the F$_{r}$-F$_{IR}$ and
L$_{r}$-L$_{IR}$ planes (Fig.~\ref{rir}).  FR~II LIGs nuclei also show a trend
in both planes, even if they cover a smaller range in radio core flux (2 dex
for the NIR detected nuclei) with respect to FR~Is. Conversely, there is no
evident link between radio and NIR emission for HIGs and BLOs.

In order to assess the presence of correlations, we performed a statistical
analysis using the Astronomy Survival Analysis (ASURV) package
\citep{lavalley92}. This package is available under IRAF/STSDAS and provides
us a tool to deal with censored data (\citealt{feigelson85},
\citealt{isobe86}). Note that the method we use is based on the
assumption that the censored data are uniformly distributed. In order to
determine whether this approach is viable, we apply the censoring analysis
only when the above assumption is satisfied. We initially check the location
of the non-detected data points with respect to the entire range of available
data.

We used the ''schmittbin'' task \citep{schmitt85} to calculate the associated
linear regression coefficients for two set of variables. Operatively, we
carried out this procedure twice, obtaining two linear regressions: firstly
considering the former quantity as the independent variable and the latter as
the dependent one, secondly switching the roles of the variables. The best fit
is represented by the bisector of these two regression lines. This followed
the suggestion of \citet{isobe90} that consider such a method preferable for
problems that require a symmetrical treatment of the two variables. In order
to estimate the quality of the linear regression, for small data sets (N $<$
30) we also derived the generalized Kendall's $\tau$ \citep{kendall83} between
the two variables, using the ``bhkmethod'' task. Otherwise, we used Spearman's
rank order correlation coefficient, using the ``spearman'' task
\citep{akritas89} only for large data sets (N $>$ 30, for the cases of HIGs
and the whole sample). The statistical parameters of all the linear
regressions are reported in Table~\ref{statist1}.

The trend shown by FR~Is (Fig.~\ref{rir}, empty yellow circle points) in the
F$_{r}$-F$_{IR}$ plane corresponds to a linear correlation
coefficient\footnote{This is derived using the VLA measurements for the radio
cores. If we consider the radio cores fluxes from VLBI, we obtain a similar
correlation, statistically indistinguishable from that obtained using the VLA
cores. This is because the core fluxes measured with VLA and VLBI differ only
marginally (see the Appendix).} of r = 0.79. The probability to obtain this
level of correlation from two independent variables is P =
2.2$\times$10$^{-5}$. The statistical parameters are even better in the
radio-infrared luminosity plane, where the linear correlation coefficient is r
= 0.85, with an associated probability P = 8.9$\times$10$^{-7}$. The
dispersions of these correlations are $\sim$0.51 and $\sim$0.48 dex, for
fluxes and luminosities, respectively. The generalized Kendall's $\tau$
coefficient, which allows us to include censored data, is higher for the
luminosities ($\tau$ = 1.18) than that for the fluxes ($\tau$ = 1.00). The
associated probabilities of no correlation, are 0.0002 and $<$0.0001,
respectively. Note that the slopes of both linear regressions are consistent
with unity.

We now turn our attention to the possible presence of correlations for FR~II
galaxies.  FR~II LIGs show a linear correlation in both planes of
Fig.~\ref{rir}.  Considering only the detected nuclei, the linear correlation
coefficient is r = 0.89 in the F$_{r}$-F$_{IR}$ plane, with a probability of a
no correlation of P = 1.0$\times$10$^{-4}$.  The censoring statistics gives a
generalized Kendall's coefficient $\tau$ = 0.77 with a probability, P =
0.0012, that a fortuitous correlation appears at a level measured by our
test statistic.  In L$_{r}$-L$_{IR}$ plane for FR~II LIGs a
linear relation (dotted line in Fig.~\ref{rir}, right panel) is also
statistically supported: the linear correlation coefficient is r = 0.93 (P =
1.1$\times$10$^{-5}$).  The radio-infrared core luminosity relation for FR~II
LIGs has a slope of 0.9$\pm$0.3.

\begin{figure}
\includegraphics[scale=0.4]{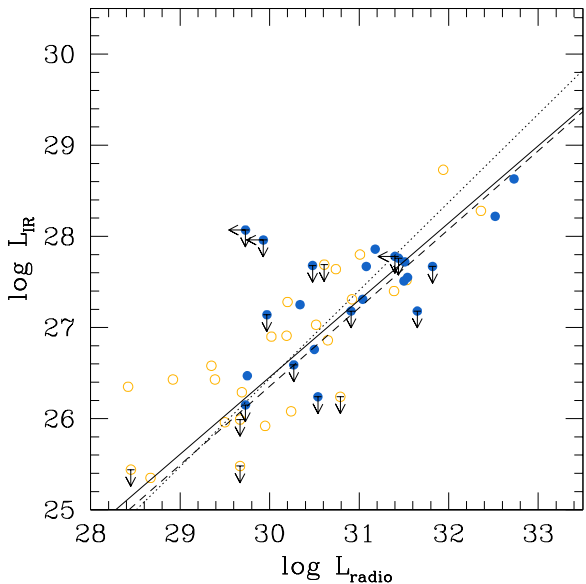}
\caption{Radio core luminosity vs. near infrared nuclear luminosity (in units
  of erg s$^{-1}$ Hz$^{-1}$) for FR~Is (empty yellow points) and FR~II LIGs
  (blue points). We show the linear correlations: the solid line is obtained
  for FR~Is and FR~II LIGs, the dotted line for FR~Is only, and the dashed
  line for FR~II LIGs only.}
\label{pippo}
\end{figure}

Therefore, an important result of our analysis is that the correlations found
for FR~Is and FR~II LIGs nuclei are consistent with each other within the
errors. This result is shown in Fig.~\ref{pippo} where for more clarity we
have marked these two classes with different symbols.

For HIGs and BLOs the absence of a trend is confirmed by the results of the
statistical analysis and in particular by the high probabilities that their
distribution in the L$_{r}$-L$_{IR}$ plane is random (see the
statistical parameters in Tab.~\ref{statist1}).

\subsection{Spectral indices}

\begin{figure*}
\includegraphics[scale=0.45]{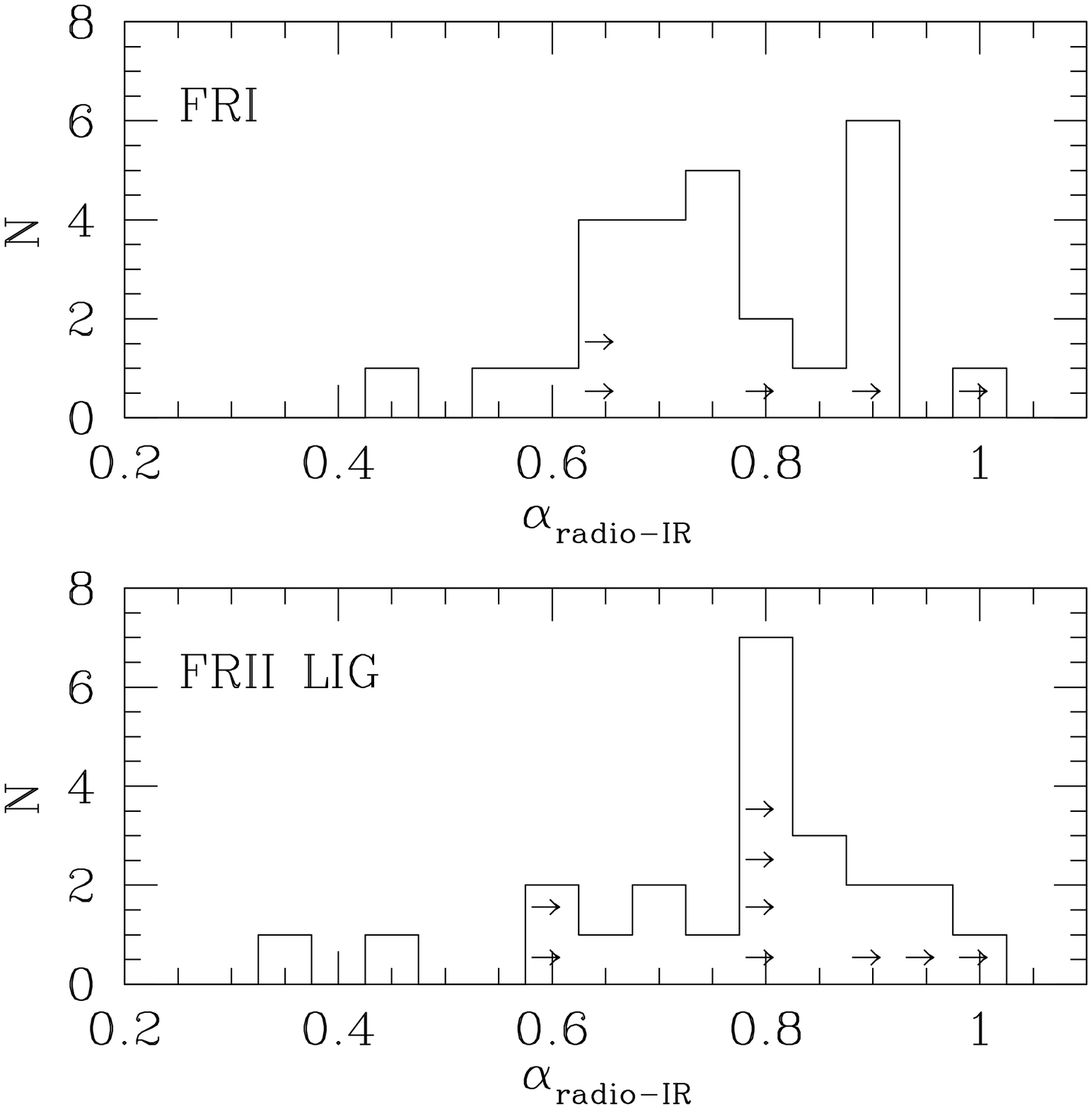}
\includegraphics[scale=0.45]{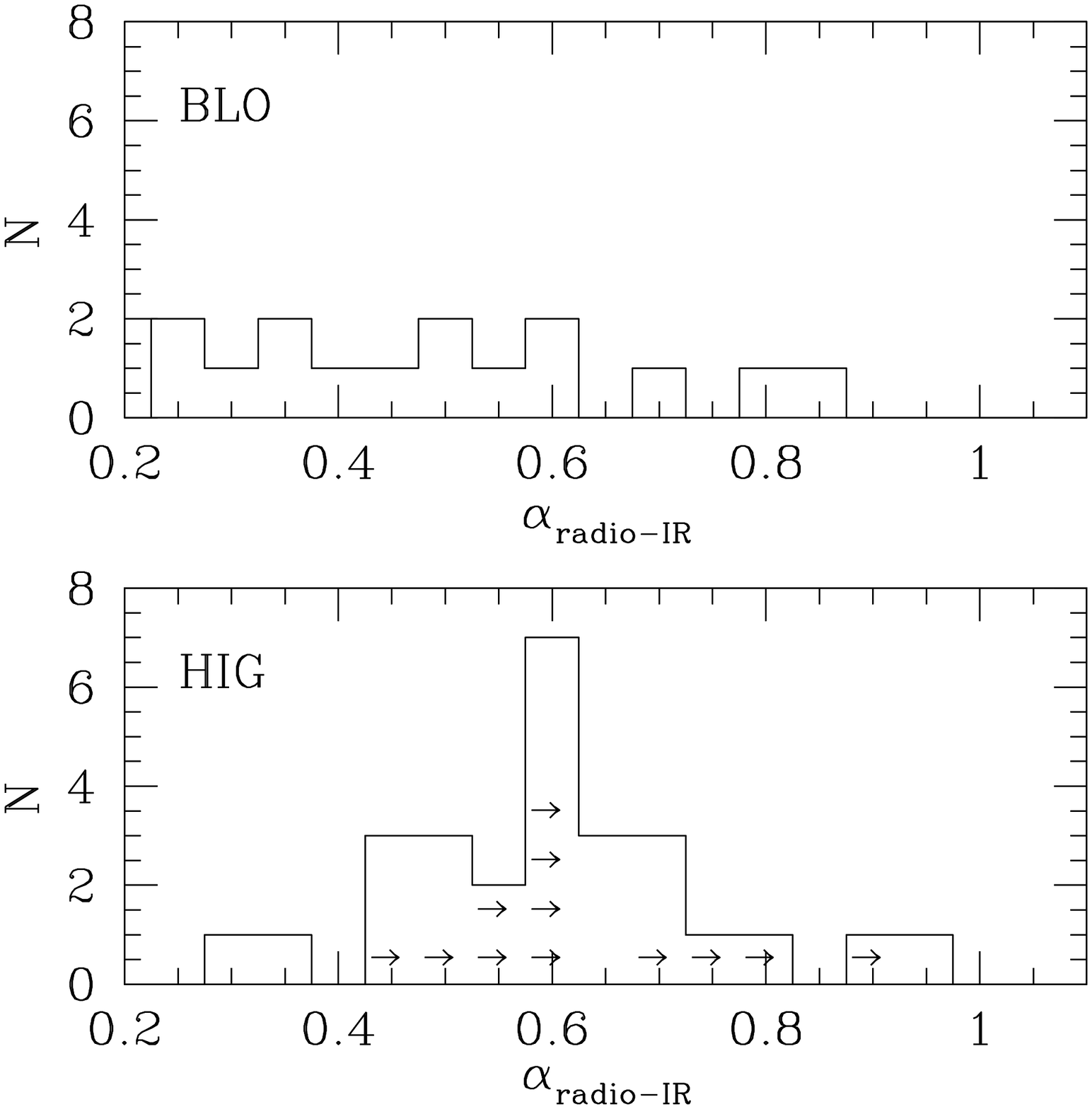}
\caption{Histograms of radio-infrared spectral indices $\alpha_{r-IR}$ for
each sub-class of the sample. The arrows represent the lower limits data.}
\label{alfaisto}
\end{figure*}

In order to derive the multi-wavelength properties of the nuclei, it is useful
to analyze the broad-band radio-infrared spectral index $\alpha_{r-IR}$.  In
Fig.~\ref{alfaisto} we show the empirical distribution in $\alpha_{r-IR}$ for
the four different classes (FR~I, FR~II LIG, HIG and BLO).  Note that the
average value of the radio-IR spectral index distributions shifts towards
lower values from FR~II LIGs to the BLOs, with FRIs and HIGs having
intermediate values.  This indicates that with respect to the IR flux, the
FR~II LIGs are the most radio-powerful objects. Another intriguing aspect
worth noting is that even though FR~II LIGs and FR~Is behave similarly in the
radio-IR plane, as they define the same correlations, their distribution of
$\alpha_{r-IR}$ are slightly different.  In fact, FR~II LIGs have on average a
slightly higher $\alpha_{r-IR}$ than the FR~I galaxies. On the other hand,
HIGs and BLOs show an near infrared excess with respect to the radio emission
and therefore their $\alpha_{r-IR}$ spectral indices are on average lower than
for the other two classes.

To assess the presence of any significant differences between the spectral
index distribution of the different classes, we can make use of the 'twosampt'
task, available in the ASURV package. The task computes several nonparametric
two-sample tests, giving a variety of ways to test whether two censored
samples are drawn from the same parent population.

We utilize the Peto-Prentice statistic \citep{peto72}, which is a
generalization of standard test for uncensored data, such as the Wilcoxon
\citep{wilcoxon45} and Kolmogorov-Smirnov \citep{chakravarti67} test. This
statistical test quantifies a distance between the empirical distribution
functions of two samples. The null distribution of this statistic is
calculated under the null hypothesis that the samples are drawn from the same
distribution.

We statistically measure the different distributions of spectral indices
between FR~Is and FR~II LIGs and between HIGs and BLOs. We can assert
that FR~Is and FR~II LIGs are not drawn from the same parent population at the
97\% confidence level. The same holds for the HIG and BLO classes at the 99\%
confidence level.

\subsection{Comparison with FR~Is' radio-optical correlation}
\label{compareC}

\citet{chiaberge99} measured the optical nuclear fluxes for most FR~I radio
galaxies in the 3CR catalog.  The optical detection rate is similar to that
found for the FR~Is of our sample. For the sub-sample of objects for which the
total radio luminosity at 178 MHz is lower than the FR~I/FR~II break
(corresponding to L$_{178\,MHz}$ $\sim$ 2$\times$10$^{33}$ erg s$^{-1}$
Hz$^{-1}$), they found a tight linear correlation between the optical and
radio nuclear fluxes with a dispersion of $\sim$0.4 dex.  In order to compare
the radio-infrared and radio-optical correlations we selected the 15 FR~Is in
common with the sample of \citet{chiaberge99} (namely 3C~29, 3C~31, 3C~66B,
3C~83.1, 3C~84, 3C~264, 3C~270, 3C~272.1, 3C~274, 3C~296, 3C~317, 3C~338,
3C~442, 3C~449, and 3C~465)\footnote{Although 3C~277.3 satisfies the selection
criteria, we excluded the object from this sample since from an careful
inspection of its radio maps we revised its radio-morphological classification
from FR~I to FR~II.}.  We estimate the linear regression between radio and
near infrared fluxes for this sub-sample: the correlation coefficient is r =
0.82 (P = 2.2$\times$10$^{-4}$) and the dispersion is $\sim$0.49 dex.  We
tested that the dispersions of the two relations, F$_{r}$-F$_{IR}$ and
F$_{r}$-F$_{o}$, are not statistically different at a significance level of
95\%.

\begin{table*}
\begin{center}
\caption{Summary of the statistical analysis of the linear regressions.}
\begin{tabular}{cccccccccccccc}
\tableline\tableline
Relation & Class & N$_{d}$ & N$_{c}$  & r$_{d}$  & P$_{r_{d}}$  & $\rho_{d}$ or $\tau_{d}$ & P$_{\rho_{d}}$ or P$_{\tau_{d}}$ & $\rho$ or $\tau$ & P$_{\rho}$ or P$_{\tau}$ & rms$_{x}$ & rms$_{y}$ & m & q \\
\tableline
F$_{r}$-F$_{IR}$ & FR~I & 21 & 5 & 0.788 & 2.2$\times$10$^{-5}$ & 1.181 & 0.0002 & 0.997 & 0.0002 & 0.477 & 0.543 & 1.0$\pm$0.2& -3$\pm$5\\
F$_{r}$-F$_{IR}$ & FR~I $^{a}$ & 15 & 0 & 0.834 & 0.00011  & 1.238  & 0.0013 & / & / & 0.442 & 0.542 & 0.8$\pm$0.2 & -8$\pm$6\\
F$_{r}$-F$_{IR}$ & FR~I $^{b}$ & 15 & 1 & 0.855 & 4.9$\times$10$^{-5}$ & 1.257 & 0.0011 & 1.233 & 0.0006 & 0.394 & 0.407 & 1.0$\pm$0.3  & -3$\pm$6\\
F$_{vlbi}$-F$_{IR}$ & FR~I   & 15 & 1 & 0.866 & 3.0$\times$10$^{-5}$ & 1.239 & 0.0012 & 1.217 & 0.0007 & 0.380 & 0.291 & 1.3$\pm$0.3 & 5$\pm$5\\
F$_{r}$-F$_{IR}$ & LIG & 11 & 12 & 0.890 & 0.00010 & 1.490 & 0.0014 & 0.775 & 0.0012 & 0.168 & 0.298 &  0.6$\pm$0.3 & -14$\pm$15\\
F$_{r}$-F$_{IR}$ &FR~I+LIG & 32 & 17 & 0.781 & 4.1$\times$10$^{-8}$ & 0.743 & $<$0.0001 & 0.645 & $<$0.0001 & 0.445 & 0.511 & 0.9$\pm$0.2  & -7$\pm$14\\
F$_{r}$-F$_{IR}$ & HIG & 15 & 12& 0.150 & 0.59 & -0.095& 0.804 & 0.296 & 0.218 &  0.542 & 0.544 & 1.2$\pm$0.4 & 3$\pm$24\\
F$_{r}$-F$_{IR}$ & BLO & 15 & 0 & 0.348 & 0.20 & 0.514 & 0.181 & / & / &  0.680 & 0.727 & 1.0$\pm$0.4 & -3$\pm$9\\
L$_{r}$-L$_{IR}$ & FR~I & 21 & 5 & 0.854 & 8.9$\times$10$^{-7}$ & 1.324 & $<$0.0001 & 1.077 & $<$0.0001 & 0.420 & 0.532 &1.0$\pm$0.2 & -3$\pm$6\\
L$_{r}$-L$_{IR}$ & FR~I$^{b}$ & 15 & 1 & 0.898 & 5.4$\times$10$^{-6}$ & 1.276 & 0.0009 & 1.167 & 0.0008 & 0.392 & 0.417 & 1.0$\pm$0.2  & -3$\pm$5\\
L$_{vlbi}$-L$_{IR}$ & FR~I  & 15 & 1 & 0.903 & 4.1$\times$10$^{-6}$ & 1.333 & 0.0005 & 1.183 & 0.0007 & 0.384 & 0.347 & 1.1$\pm$0.2 & -7$\pm$5\\
L$_{r}$-L$_{IR}$ & LIG & 11 & 12 & 0.931 & 1.1$\times$10$^{-5}$ & 1.491 & 0.0014 & 0.696 & 0.0011 & 0.211 & 0.306 & 0.9$\pm$0.3 & 0$\pm$12\\
L$_{r}$-L$_{IR}$ & FR~I+LIG & 32 & 17 & 0.890 & 1.8$\times$10$^{-8}$ & 0.900 & $<$0.0001 & 0.688 &$<$0.0001 & 0.363 & 0.487 & 0.8$\pm$0.2 & 1$\pm$18\\
L$_{r}$-L$_{IR}$ & HIG & 15 & 12 & 0.503 & 0.056 & 0.514 & 0.181 & 0.461 & 0.054 & 0.637 & 0.606 & 1.3$\pm$0.5 & -13$\pm$27\\
L$_{r}$-L$_{IR}$ & BLO & 15 & 0 & 0.2398 & 0.39 & 0.152 & 0.692 & / & / & 0.488 & 0.802 & 0.8$\pm$0.5 & 4$\pm$14\\
\tableline

\end{tabular}
\tablecomments{Column description: (1) relation studied; (2) class of
radio-galaxies considered for the relation; (3) number of detected nuclei
included in the relation considering both the quantities; (4) number of
censored nuclei included in the relation considering both the quantities; (5)
r$_{d}$ linear correlation coefficient for the relation obtained only with the
detected data; (6) P$_{r_{d}}$ probability associated with r$_{d}$ that the
correlation is not present; (7) $\rho_{d}$ Spearman's rank coefficient (if
N$_{d}$ $>$ 30) or $\tau_{d}$, generalized Kendall's tau (if N$_{d}$ $<$ 30),
for the relation obtained only with the detected data; (8) P$_{\rho_{d}}$ (or
P$_{\tau_{d}}$) probability associated with $\rho_{d}$ (or $\tau_{d}$) that the
correlation is not present; (9) $\rho$ Spearman's rank coefficient (if
N$_{d+c}$ $>$ 30) or $\tau$, generalized Kendall's tau (if N$_{d+c}$ $<$ 30),
considering all the data; (10) P$_{\rho}$ (or P$_{\tau}$) probability
associated with $\rho$ (or $\tau$) that the correlation is not present; (11)
rms$_{x}$, rms on x-axis for the relation obtained only with the detected data;
(12) rms$_{y}$, rms on y-axis for the relation obtained only with the detected
data; (13)-(14) slope coefficient m and intercept coefficient q for the linear
regression (y = m*x + q) obtained considering all the data. In the rows marked
with $a$ the considered sample is composed of the FR~I objects with L$_{178
MHz} <$ 2$\times$10$^{26}$ W Hz$^{-1}$ shared with sample studied by
\citet{chiaberge99}. In the rows marked with $b$ the considered sample is
composed of the FR~Is whose the VLBI radio core measurement is available. With
LIG we consider FR~II LIG.}
\label{statist1}
\end{center}
\end{table*}

\section{Discussion}
\label{discussion}

In the following we consider the properties of 
each group of 3CR radio galaxies separately.

\subsection{FR~I galaxies}

The presence of a linear correlation between FR~I radio and NIR cores, both in
flux and luminosity, argues for a real physical connection between the two
variables. This is indeed not the result of a common dependence on redshift of
the two quantities, as shown by the fact that the statistical parameters of
both correlations are similar.  Following \citet{chiaberge99}, that
interpreted the optical-radio core correlation as the indication that
synchrotron emission dominates the nuclear emission at both wavelengths, we
argue that the correlations between the radio and the NIR nuclei described
here can be explained by the same physical mechanism. More precisely, the
presence of such a tight correlation between the NIR and radio cores indicates
that non thermal (synchrotron) radiation dominates their emission at all
considered wavelengths, including the X-ray as shown by
\citet{balmaverde06}. \citet{capetti07} found that the optical HST nuclei
of 9 3CR/FR~I, which also belongs to our sample, are highly polarized. This
strengthens the synchrotron origin of their nuclear emission.

The correlations among NIR, VLA and VLBI cores and the fact that the fluxes of
the VLA and VLBI cores (see Appendix) are essentially equal suggest that the
NIR cores are typically produced on scales consistent with or even smaller
than the size of the VLBI cores. In fact, according to various jet models
\citep[e.g.,][]{ghisellini85}, it is likely that the radio and the IR/Optical
radiation is produced in the innermost region of the jet, close to the black
hole ($\lesssim$ 10$^{16}$ cm), while the radio is emitted on larger scales.

The dominance of synchrotron radiation even in the NIR indirectly confirms
that accretion in most FR~I nuclei occurs at low rates and/or with low
radiative efficiency (e.g., \citealt{baum95}, \citealt{chiaberge99},
\citealt{allen06}, \citealt{balmaverde08}). However, there are exceptions
to this result: the clearest example is possibly 3C~120 (a radio galaxy
which is formally not part of 3CR sample) which displays a radio morphology
typical of FR~Is (e.g. \citealt{walker01}) while its nuclear emission shows
the clear signature of broad lines and radiatively efficient accretion
\citep{phillips75}. On the other hand, exceptions are also present among
BL~Lacs, the putative ``beamed'' counterpart of LIG population. A number of
BL~Lacs also show (faint) narrow and/or broad emission lines
(e.g. \citealt{vermeulen95}) and others have FR~II-like extended radio
emission (e.g. \citealt{antonucci86,kharb10}).

\citet{chiaberge99} pointed out that the dispersion of the radio-optical
correlation ($\sim$ 0.4 dex) is consistent with dust absorption, provided that
the $A_V$ is randomly distributed among the sources and does not exceed $\sim$
2 mag. Since the extinction in H-band is 4.5 times lower than in the optical
R-band, the observations presented in this work are far less sensitive to the
presence of dust along the line of sight than those in \citet{chiaberge99}.
Therefore, one would expect the dispersion of the infrared-radio correlation
to be smaller than that of the radio-optical, if that is due to absorption.
Our results instead show that the rms of the two correlations are similar.
Thus, the most likely explanation for the scatter is that variability of the
nuclei plays a fundamental role, while the effect of absorption is less
significant. Indeed, this is supported by the variability observed for the
nuclear flux of 3C~317 \citep{chiaberge02b} and of 3C~274
(\citealt{tsvetanov98}, \citealt{perlman03}).  Conversely, the presence of
dust appears to play a role in dimming the nuclear flux on the UV
\citep{chiaberge02b}. Note that data in the literature indicate that the
radio core variability is typically at 10\% over a time scale of several years
\citep{hine80,ekers83}. Therefore the main source of the scatter in the
radio-NIR correlation is mostly likely due to a variable NIR nuclear
component.

Furthermore, the region of the jet base that produces the optical/NIR
emission in FR~Is is likely to be mildly relativistic (e.g.,
\citealt{chiaberge99,xu99a}). Therefore, we  consider the effects of
Doppler boosting on our results. Firstly, the range in NIR nuclear luminosity
might be enhanced by relativistic beaming, which would boost (or de-boost) the
emission of FR~I nuclei, depending on the viewing angle.  Objects seen under a
small angle between the line-of-sight and the jet axis would appear boosted,
while objects observed closer to the plane of the sky would be de-boosted.
Secondly, nuclear variability might be enhanced by relativistic beaming when
the flux variation is associated with 1) a change in the jet bulk Lorentz
factor, or 2) a change in the jet direction, which would result in a different
angle between the jet direction and the line-of-sight.

However, as noted by \citet{chiaberge99}, the bulk Lorentz factor of the
component responsible for the nuclear optical/NIR emission is most likely
around $\Gamma \sim 2$ and the source is possibly a ``slower'' shear layer in
the  jet.  Therefore, the effects of relativistic
beaming should not dramatically affect our results.

Are the NIR data telling us something new about the presence of optically
  and geometrically thick obscuring tori in FR~Is with respect to the optical
  observations? The results derived considering the UV, optical, and NIR bands
  are only apparently in contradiction. In fact, the median absorption toward
  FR~I nuclei derived from UV observations is $A_V \sim 1.3$. This corresponds
  to only 0.2 mag in H band, negligible with respect to a dispersion of
  $\sim$0.5 dex, and to $\sim 3$ mag at 2500 \AA.

The detection rate of FR~I nuclei in the NIR (81\%) is similar to that found
in the R band by \citet{chiaberge99}. All sources with an optical nucleus also
show a NIR counterpart, except 3C~310. Instead, in 3C~75N a faint dust lane
prevents the study of its nuclear regions in the optical, while its NIR core
is clearly detected. Furthermore, the absence of FR~I outliers in the
L$_{IR}$/L$_{O}$ ratio (see right panel of Fig.~\ref{oir}) can be interpreted
as a hint that the {\sl detected} FR~I nuclei are not affected by significant
absorption ($A{_V} \lesssim 2$ mag).

\begin{figure}
\includegraphics[scale=0.45]{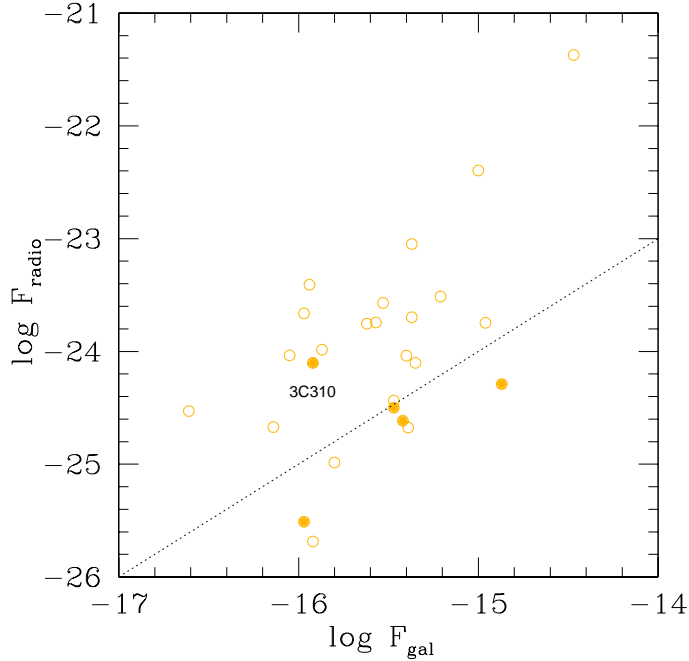}
\caption{Radio core flux density, log F$_{radio}$ (erg s$^{-1}$ cm$^{-2}$
  Hz$^{-1}$), versus central (on a circular region of r=0$\farcs$15) surface
  brightness of the host galaxy, log F$_{gal}$ (erg s$^{-1}$ cm$^{-2}$
  \AA$^{-1}$ arcsec$^{-2}$) for all FR~I. Objects with detected near infrared
  nuclei are marked with empty yellow circles, while those with non-detection
  by filled yellow circles. The dotted line marks a constant ratio between the
  radio core and galaxy emission.}
\label{contrast}
\end{figure}

The reason for the non-detection of some of the FR~I NIR nuclei might instead
be due to a small contrast between the nucleus and the host galaxy stellar
emission. Assuming that the radio and the near infrared emission of FR~I
nuclei are intrinsically tightly correlated, the intensity of their radio
cores provides a robust prediction of their infrared nuclear flux. Following
the analysis on the B2 radio-galaxies by \citet{capetti02}, we plot the radio
core fluxes of all FR~Is against the galaxies central surface brightness
(Fig.~\ref{contrast}). Four out of the five undetected nuclei are indeed among
those with the lowest predicted contrast against the galaxy emission. 3C~310
is the only object that does not have an NIR detected nucleus and apparently
shows an rather larger contrast with galaxy emission than that of the other
four non-detected nuclei. However, since its optical nucleus is detected, its
NIR non-detection is likely not due to a obscuration.

Summarizing, we argue that in most cases we have a direct view of FR~I nuclei
and that the few undetected NIR central sources are below the detection
threshold set by the contrast with the underlying host galaxy emission. Note
that the detection of massive dusty structures in the central regions of some
FR~Is (e.g. \citealt{okuda05}, \citealt{das05}) does not contrast with our
result that the NIR nuclei are mostly unobscured. In fact dust is distributed
in large kpc-scale geometrically thin disks which do not appear to `hide' the
nuclei to our line-of-sight.

\subsection{FR~II: Low Ionization Galaxies LIG}

\citet{chiaberge02} found that the nuclei of FR~II LIGs lie on the FRI
radio-optical correlation. The sample of FR~II LIGs available in this work is
substantially larger than the sample those authors considered, both because of
the increasing number of FR~II LIG identifications and of the larger sample
size. This allows us to establish the presence of a correlation between the
radio and the NIR emission within the sample of FR~II LIGs itself. An
important result is that the correlations for FR~Is and FR~II LIGs are
statistically indistinguishable. This strengthens the hypothesis that the LIGs
are FR~II galaxies with FR~I-like nuclei. Such a nuclear similarity implies
that non-thermal synchrotron radiation (e.g., from a relativistic jet)
dominates their nuclear emission in NIR band and that accretion may occur with
a radiatively inefficient process and/or at low accretion rates in FR~II LIGs
as well.

The large fraction of FR~II LIGs in our sample ($\sim$33\% of the FR~IIs)
proves that these objects are not rare exceptions in the FR~II class.  This
class of objects show nuclear properties similar to those of FR~Is, whereas
their radio morphologies span the range from classical double to X-shaped, and
from compact source to Fat Double (FD). The FD morphology \citep{owen89}
consists of a FR~II radio structure with elongated and diffuse lobes and
sometimes with prominent jets, and may be considered as an intermediate class
between FR~Is and FR~IIs.  At least half of the FR~II LIG sample show a FD
radio morphology.  The large variety of extended radio morphologies associated
with a central LIG-like AGN makes it difficult to understand how the
properties of the core and extended radio structures are related
(e.g. \citealt{bicknell84,bicknell94,gopal01}). The existence of FR~II LIGs as
a class shows that radiatively inefficient accretion disks are able to produce
jets that give rise to the typical large scale FR~II morphology. Another
possibility is that FR~II LIGs constitute an evolutionary (or transient) state
of the life of the FR~II class.  However, the environment of FR~II LIGs is
different from that of other classes of FR~IIs, as FR~II LIGs are
predominantly found in dense groups or clusters
\citep{chiaberge00,hardcastle04}. That seems to rule out the evolutionary
scenario.

The lower detection rate of FR~II LIG nuclei with respect to that of FR~Is is
most likely due to the reduced contrast between their faint IR nuclei and the
surrounding stellar emission of the host galaxy, because of the higher
redshift of FR~II LIGs as compared to that of FR~Is.  In fact, the fraction of
detected FR~II LIG nuclei decreases as redshift increases, from 55 \% for
z$<$0.15 to 29 \% for z$>$0.15.  However, at this stage we cannot rule out
that the presence of significant obscuration along the line of sight to the
FR~II LIGs with non-detected nuclei might also contribute to the reduce the
fraction of detected FR~II LIG nuclei. The best way to probe obscuration in
FR~II LIGs is through X-ray spectral analysis.  FR~II LIGs indeed seem to lack
significant absorbing column densities, $N_{\rm H} \sim 10^{23}$ cm$^{-2}$
\citep{hardcastle06}.

The fact that the central engines of FR~II LIGs and FR~Is are similar
(i.e., they are both unobscured and lack a significant broad line region) has
important bearings for the unification schemes.  As pointed out by
\citet{laing94}, the broad distribution in core-dominance is consistent with
FR~II LIGs being a randomly oriented sample. Since FR~II LIGs also appear to
lack significant broad emission lines, when observed along the jet axis, these
objects would appear as BL~Lacs, in agreement with 
\citet{chiaberge00b} in light of their findings in the optical band. Such a
unification scenario, which was previously explored by \citet{jackson99} based on the
optical spectral properties of FR~II LIGs, is in apparent contrast with zeroth
order unification scheme which associates all FR~II with quasars.  However,
besides being in agreement with our results, it may also account for the
presence of BL Lacs with an FR~II morphology.

\begin{figure}
\includegraphics[scale=0.4]{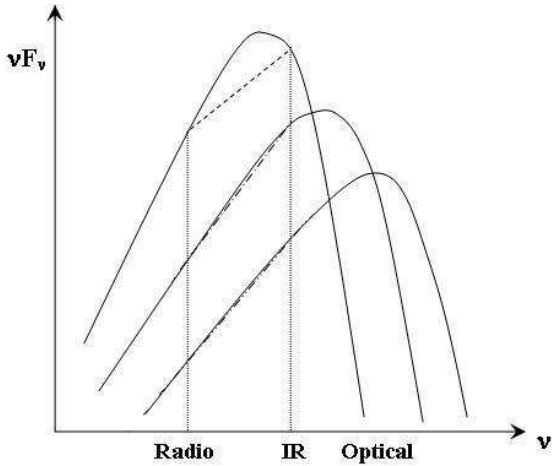}
\caption{The Spectral energy distributions (SED) of three blazars in order to
  explain the so-called scenario \citep{fossati98}. In the three blazars, the
  synchrotron emission peak shifts towards lower frequencies as the radio
  luminosity at 5 GHz increases, which also corresponds to a higher luminosity
  of the synchrotron emission peak. The dashed lines represent the
  radio-infrared spectral indeces of each SED. As the peak in the SED
shifts towards lower frequencies, $\alpha_{r-IR}$ steepens.}
\label{sed}
\end{figure}

Nevertheless, FR~II LIGs do show some differences with respect to FR~Is. In
fact, as shown in Fig.~\ref{alfaisto}, FR~II LIGs and FR~Is have slightly
different radio-IR broad-band spectral indices. If FR~II LIGs are indeed the
counterparts of BL~Lacs with FR~II radio morphology, the spectral index
discrepancy can be explained in the framework of the so-called ``blazar
sequence'' \citep{fossati98}. Such a scenario predicts a smooth transition in
the properties of blazars' spectral energy distribution (SED).  In blazars,
the synchrotron emission peak is observed to shift towards lower frequencies
as the radio luminosity at 5 GHz increases. The luminosity of the synchrotron
emission peak also increases following the same trend.  This gives rise to a
sequence having high energy peaked BL~Lacs at the lowest luminosities, low
energy peaked BL~Lacs at intermediate luminosities, and flat spectrum radio
quasars at the highest bolometric powers.  In Fig.~\ref{sed} we schematically
show what the effect of such a scenario on the broad band spectral indices
would be.  As the peak in the SED shifts towards lower frequencies,
$\alpha_{r-IR}$ steepens.  Therefore, the steeper spectral indices we found in
FR~II LIGs might be explained in terms of a slightly higher 5GHz core
luminosity with respect to that of FR~Is.  FR~II LIGs would thus correspond to
intermediate power blazars such as the low-energy peaked BL~Lacs.

The possibility that, conversely, the change in $\alpha_{r-IR}$ is due to the
redshift of the SED for the more distant objects can be excluded using the
formula presented by \citet{trussoni03}. The formula, originally derived for
estimating the effect of relativistic beaming, can be used to to account for
the effect of redshift on the broad-band spectral indices, by substituting
R$_{\delta}$, the ratio of the beaming factors, with $(1 + z)$.  The variation
in the radio-infrared spectral due to the frequency shift for an object at
z$\sim$0 moved to z=0.3 is only $\Delta\alpha_{r-IR} \sim 0.02$, a correction
that has a negligible effect on the results.

\subsection{FR~II: Broad Line Objects BLO}
\label{discblo}

According to the standard unification scheme, the BLOs are objects where the
nuclear region is unobscured along our line of sight
\citep[e.g.][]{barthel89,antonucci85}. This idea is supported by the fact that
all the NIR nuclei are detected for BLOs.

The BLOs show a strong NIR excess, of up to two orders of magnitude in
luminosity, with respect to the correlation defined by the FR~I (and FR~II
LIG) radio-infrared nuclei (Fig.~\ref{rir}, right panel). A similar excess was
already found in the optical band by \citet{chiaberge02} interpreted that as
thermal emission from the accretion disk.

Let us now explore the origin of their NIR nuclear light.  The
infrared-optical spectral index for BLOs covers the range $\alpha_{IR-o}$
$\sim$ 0 - 2 (with the exception of 3C~111 for which $\alpha_{IR-o}$ = 3.3).
Such a large range might result from a wide distribution of spectral
indices (see Fig.~11 in \citealt{elvis94}). In addition, variability may
smear the intrinsic distribution. Therefore we consider the median value
$\alpha_{IR-o} = 0.95$.  This value indicates that the Spectral Energy
Distribution (SED) of BLOs is essentially flat in a $\nu$L$_\nu$
representation.

To better investigate the origin of NIR emission in the BLOs of our sample we
compare their properties with those of the radio-loud QSOs (RLQSO) from
\citet{elvis94}. As demonstrated by those authors, the averaged SED of QSOs
shows a minimum at $\sim$1$\mu$m.  For wavelengths shorter than 1$\mu$m the
light is likely to be dominated by the accretion disk (i.e., the big blue
bump), while for longer wavelengths a broad peak is observed, which is
interpreted as emission from hot dust. The flat spectral index of BLOs is an
indication that, on average, the optical and NIR HST observations are located
on opposite sides of the SED minimum.  Therefore, by analogy with RLQSOs, we
then conclude that in most 3CR/BLOs the H band light is mainly produced by the
high temperature tail of the hot dust emission, located at the inner face of
the obscuring material.

We estimate a spectral index of $\alpha_{IR-o}$ = 0.35 from the averaged SED
of RLQSOs after adopting the mean redshift of BLOs (z = 0.14). The difference
of the spectral indices between RLQSOs and BLOs can be accounted by an
NIR/optical flux ratio of the BLOs larger by a factor $\sim$2 than that of
RLQSOs. This is an indication of a relatively larger contribution from the
dust in infrared emission of BLOs than that observed in RLQSOs. Furthermore,
reminding the reader that the RLQSO have optical and NIR luminosities a factor
of 100-1000 higher than those of the 3CR/BLOs, there is a dependence of the
optical-infrared spectral index on the luminosity of the object.

\subsection{FR~II: High Ionization Galaxies HIG}
\label{discusHIG}

At odds with the results found for FR~II LIGs and FR~Is, the NIR
emission of HIG nuclei does not correlate with the radio cores. Furthermore,
most of the detected nuclei exceeds the emission of FR~Is and FR~II LIGs at a given
radio-core luminosity, suggesting an additional component dominating over the
jet emission, similar to what we found for the BLOs.

However, HIG nuclei are substantially fainter than those of the BLOs, on
average by a factor of 50.  This is expected in a standard unification scheme,
since the HIGs are considered to be the obscured counter parts of BLOs.  This
result indicates that substantial obscuration is still present in the 
NIR, despite the reduced effects of dust in the H band.

\begin{figure}
\includegraphics[scale=0.45]{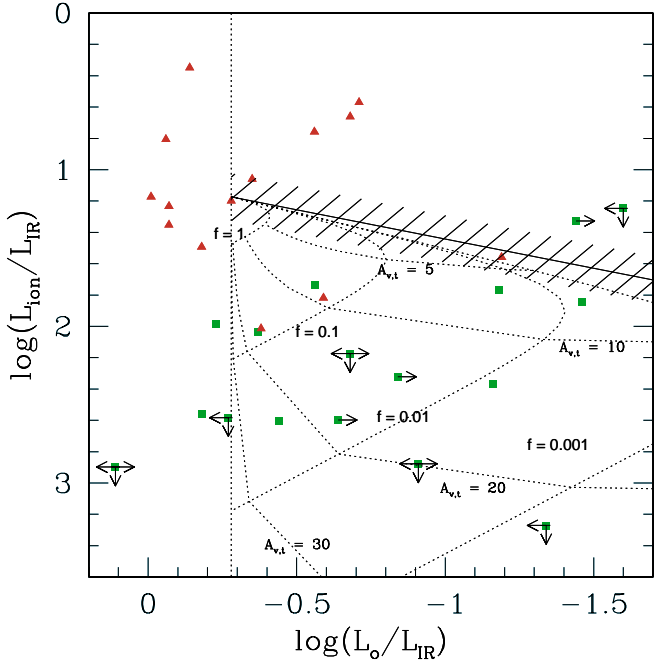}
\caption{Ratio between the ionization luminosity L$_{ion}$ and $\nu$L$_{IR}$
  versus the ratio between L$_{o}$ and L$_{IR}$. The BLOs are the red
  triangles and the HIGs the green squares. The solid oblique line is the
  'reddening line' that purely obscured objects should follow. The origin of
  this line is determined by the ratios measured for BLOs, i.e.
  log(L$_{ion}$/$\nu$L$_{IR}$) = 1.17$\pm$0.15 and log(L$_{O}$/L$_{IR}$) =
  -0.28$\pm$0.11. The dashed region reproduces the uncertainty of its
  origin. The dotted vertical line corresponds to the average value of
  log(L$_{O}$/L$_{IR}$) measured for BLOs. The grid of dotted curves represent
  the effects of varying the nuclear A$_{v,t}$ and the fraction of scattered
  light $f$ (see the Eq.3 in \citet{marchesini05}).}
\label{marchesini}
\end{figure}

In order to investigate the origin of NIR emission in HIGs, we follow the
method described by \citet{marchesini05} who studied a complete sub-sample of
the 3CR radio catalog, which includes all HIGs with z $<$ 0.3, observed in
K$^{\prime}$ band from the ground.  They first estimated the ionizing
luminosity, L$_{ion}$ = L$_{NLR}$ C$^{-1}$ =0.0675 $\times$ L$_{[O III]}$
(e.g. total luminosity in narrow lines, L$_{NLR}$, estimated from [O~II] and
[O~III], \citealt{rawlings91}) for HIGs and BLOs assuming the same covering
factor for the two classes, C = 0.01. Then, for each source,
\citet{marchesini05} plotted L$_{ion}$/L$_{IR}$ against L$_{o}$/L$_{IR}$,
which are both sensitive to nuclear obscuration. They found that HIGs do not
follow the `reddening line' (see caption of Fig.~\ref{marchesini}) in such a
plane, as one might expect if they were simply ``partially obscured'' BLO
nuclei. This lead them to conclude that a substantial role is played by
scattered nuclear light.

With respect to \citet{marchesini05} our analysis can rely on {\sl direct}
measurements of the NIR nuclei and these are available for a larger
number of HIG, due to better sensitivity of the HST images to faint IR nuclei.
Furthermore, we also have NIR data for the nuclei of BLO, thus providing
a more robust estimate of the location of unreddened sources.

In agreement with the results \citet{marchesini05} we found that only a few
HIGs are located along the reddening line in Fig.~\ref{marchesini} but this is
clearly not the case for the whole population.

We then include the effects of scattered nuclear light. In this scenario a
dusty torus attenuates the direct nuclear light by an extinction A$_{V,torus}$
and a fraction $f$ of the total nuclear light L$_{IR}$ is scattered into our
line of sight (see Eq. 3 and in \citealt{marchesini05}).  The effects of
varying A$_{V,torus}$ is as follows: an increase of the nuclear obscuration
simply corresponds to an increase of L$_{ion}$/$\nu$L$_{IR}$.  The behavior of
L$_{o}$/L$_{IR}$ is more complex: initially, an increase of the absorption
moves the point (starting from the average location of BLO) to the right,
along the reddening line.  At larger values of A$_{V,torus}$ the direct
(transmitted) optical light becomes fainter than the scattered component,
while the infrared nucleus is still substantially unaltered. At even higher
values of A$_{V,torus}$, the infrared transmitted component starts to
decrease; since the scattered optical component remains constant this causes
an increase of L$_{O}$/L$_{IR}$, producing a turnover in the source path.  The
track ends at the same ratio of L$_{o}$/L$_{IR}$ found for BLOs, but shifted
downward by a factor linked to A$_{V,torus}$.  The location of the turnover
instead depends on the fraction of scattered light $f$, moving to smaller
L$_{o}$/L$_{IR}$ and to larger L$_{ion}$/$\nu$L$_{IR}$ with decreasing $f$.

Because both the IR and the optical emission is most likely variable,
therefore non-simultaneous measurements may lead to large errors in the
estimate of the intrinsic L$_{O}$/L$_{IR}$. Nevertheless, this does not
compromise the analysis: while a handful HIGs can be interpreted with being
purely obscured BLOs, most of them (approximately 3/4 of the sample) require
obscuration as well as scattered light, in the range of
10$<$A$_{V,torus}$$<$30 and 0.02\%$<f<$10\% respectively. This agrees with the
general results of \citet{marchesini05} (they derived a fraction of
$\sim$70\%) and also considering HIGs on a object-by-object basis.

Our analysis of the brightness profiles of the 3CR sources always assumed that
  the nuclear NIR components are unresolved.  While for the other classes of
  3CR galaxies the nuclear sources are directly associated with the AGN (and
  thus effectively point-like at our resolution), for HIGs our results suggest
  that they are in general dominated by scattered light, a spatially resolved
  component. This is confirmed by several observations of imaging polarimetry
  and spectropolarimetry of radio-galaxies (e.g.
  \citealt{tadhunter00,ramirez09}). We then re-modeled the observed NIR
  brightness profiles taking into account the possibility that the central HIG
  sources are actually resolved and can be described by a gaussian light
  distribution.  We considered in detail the three objects
  (3C~234\footnote{3C~234 is a clear well-known case of optically obscured AGN
  with broad H$\alpha$ in polarized flux
  (\citealt{antonucci82,antonucci84,young98}).}, 3C~300, and 3C~349) that
  should have the larger scattered light fraction, being located closer to the
  grid origin in Fig.  \ref{marchesini}.  For sizes larger than $\sim$
  $0\farcs1$ significant residuals start to appear in the model fitting and we
  consider this value as an upper limit to the size of the scattering
  region. This corresponds to a range of 300-400 pc for these three objects,
  consistent with the size of the Seyfert polar scattering zone. This does not
  mean that there is no scattered light outside this region (as in
  e.g. 3C~321, \citealt{hurt99}), but it must be
  taken as an indication that the bulk of scattering occurs within this
  scale. The presence of low surface brightness scattered components (in
  contrast to the host galaxy) can only be probed with imaging polarimetry.

We conclude that HIG, unlike FR~II LIGs, are consistent with the interpretation that
they host a hidden quasar, but they are not simply partially obscured BLO.
Scattered light must provide a substantial contribution to their nuclei in
both the optical and NIR bands.

\section{Summary and Conclusions}

We have analyzed 1.6 $\mu$m near infrared images of 100 3CR radio galaxies
from HST NICMOS in the F160W band with z $<$ 0.3, $\sim$90\% of the whole 3CR
sample. On the basis of the radio morphology we divided the sample between
FR~I and FR~II and from the point of view of their optical spectra we also
classified them as Low Ionization Galaxies (LIG), High Ionization Galaxies
(HIG), and Broad Line Objects (BLO).

We measured the nuclear NIR luminosity of all the objects, by subtracting the
1-D radial brightness profile (modeled with a S\'{e}rsic or core-S\'{e}rsic
law representing the galaxy emission) from the total profile derived from the
isophote fitting of the images.

The NIR nuclear emission is found to be strongly correlated with radio cores
for FR~Is. The linear correlation extends over $\sim$ 4 dex in luminosity with
a rms of 0.4 dex and it has slope consistent with unity, as already found for
the relations with optical and X-ray nuclei. This correlation provides
further support to the identification of their nuclear emission as the
synchrotron radiation produced in the inner part of the relativistic jet. In
turn this implies a low contribution from thermal emission, an indication for
the presence of radiatively inefficient disks. The similar dispersions for the
correlations in NIR and in optical with the radio cores ascribes to the
nuclear variability the dominant role of this scatter, rather than to
obscuration.

The large number of FR~II LIGs present in this sample made possible for the
first time a detailed analysis their nuclear properties in radio-infrared
plane.  We found a strong linear correlation between radio and NIR emission.
The multiwavelength properties of FR~II LIG nuclei are statistically
indistinguishable from those of FR~Is, indicating a common structure of the
central engine, despite the difference in radio morphology and radio
power. The similar nuclear properties suggest that the nuclei of FR~II LIGs
belong to the same class of those of FR~Is, i.e., the nucleus is dominated
by non-thermal emission from a relativistic jet.

All BLOs have detected NIR nuclei and show an NIR excess with respect to the FR~I-FR~II
LIG radio-infrared correlation. This excess does not correlate with the radio
cores, suggesting the presence of an additional infrared emitting component.
In analogy with the properties of radio-quiet QSO, their optical nuclei most
likely originate from the thermal emission of the accretion disk.  Since
their optical-infrared spectral indices are essentially flat, this suggests
that thermal infrared emission from hot dust dominates in the NIR.

The origin of the NIR nuclei of HIGs is more complex. Although a few HIG
nuclei lie on the FR~I-FR~II LIG radio-infrared correlation, most of them show
a large NIR excess which does not correlate with the radio cores. This means
that synchrotron emission from the jet base is not sufficient to explain their
nuclear luminosities. Nonetheless, the HIG NIR nuclei are less luminous than
those of BLOs, by a factor of 50 on average, suggesting partial obscuration,
as expected by the unified model. Yet the infrared-optical luminosity ratios
for HIGs are not consistent with a model of HIGs as the purely-obscured
counterpart of the BLOs. We found that this inconsistency can be solved by
including a significant contribution from nuclear light reflected in a compact
circumnuclear scattering region. The sources which show the largest scattered
light contribution, have a scatter region of 300-400 pc ($\sim
0\farcs1$).

The analysis of the properties of the NIR nuclear sources of 3CR
radio-galaxies presented here broadly confirms the indications derived from
the study of the optical nuclei. In particular, the presence of two classes of
radio-loud AGN, LIG and HIG, with different nuclear properties is
strengthened.

FR~I radio-galaxies as well as the sub-population of sources with FR~II
morphology characterized by a low ionization optical spectrum belong to the
same class from the nuclear point of view. In these sources the mechanism of
emission related to the presence of non-thermal plasma from their radio-jets
dominates over the contribution from thermal radiation related to the
accretion process. Moving from optical to infrared observations has no
substantial effect despite the reduced absorption at longer observing
wavelengths. However the NIR band is not the optimal band to investigate the
issue of obscuring tori. The data preseneted here leave the results based on
optical observations unchanged, bearing in mind that for most of FR~Is we have
a clear view to their optical/infrared nucleus. While MIR Spitzer spectra
\citep{leipski09} obtained with a large slit size (10.7\arcsec) do show an
infrared excess with respect to the jet emission in three FR~I sources (and in
one FR~II LIG), our results on 3CR sample do not directly require the presence
of dusty torus on the line of sight to the FR~I nucleus. Furthermore,
although \citet{leipski09} argue that the dust emission is powered the
nucleus, the comparison between MIR spectra and HST data is extremely
difficult because of the different spatial resolution of the two
instruments. On the other hand, at this stage we cannot rule out the presence
of a dusty torus in these objects because of the higher distance of the FR~II
LIGs than that of FR~Is. Despite the lack of such information, the nuclear
properties of FR~II LIG nuclei are found to be statistically indistinguishable
from those of FR~Is. How all the sources of the FR~I/FR~II LIG group can be
associated with such a large range of radio power, while still maintaining
their disks in radiatively inefficient state, and what causes the transition
to the FR~II morphology in LIGs remain open questions.

The NIR data provide new insights also for the second class of objects, formed
by HIGs and BLOs. The BLO NIR nuclei are likely to be dominated by the
emission from hot dust located in the immediate vicinity of the active
nucleus. In HIG we found many examples of nuclei in which the NIR emission is
substantially larger than in the optical. Due to the proximity of the two
bands, the only mechanism that can explain this effect is the presence of
nuclear structures covered by absorbing material, optically thin at least in
the infrared. Scattered light must provide a substantial contribution to their
nuclei in both the optical and NIR bands. The lack of broad emission line in
HIGs is universally associated with a substantially decreased luminosity of
their nuclei in the infrared band and to a different angle of sight to the
nucleus. This implies that obscuration still plays a substantial role.

The detailed comparison of radio, optical, and infrared data for the 3CR
sample also shows that the differences in the relative contribution in these
bands, on a object-by-object basis, is dominated by variability and not by
absorption. Thus, only  simultaneous nuclear data  would allow us to explore 
intrinsic differences among the sources of the sample. The
lack of such data effectively hampers the use of the nuclear Spectral Energy
Distribution to enlighten how they vary with e.g., source power and viewing
angle.

\acknowledgments

R.D.B. acknowledges the financial support (HST-GO-11219.07-A grant) from Space
Telescope Science Institute, Baltimore. We are grateful to the referee,
R. Antonucci for the extremely careful and detailed report which stimulated us
and significantly contribuited to greatly improve the paper. We also thank
G.~K. Miley for useful comments.

\appendix
\label{appendix}

\begin{figure*}
\includegraphics[scale=0.45]{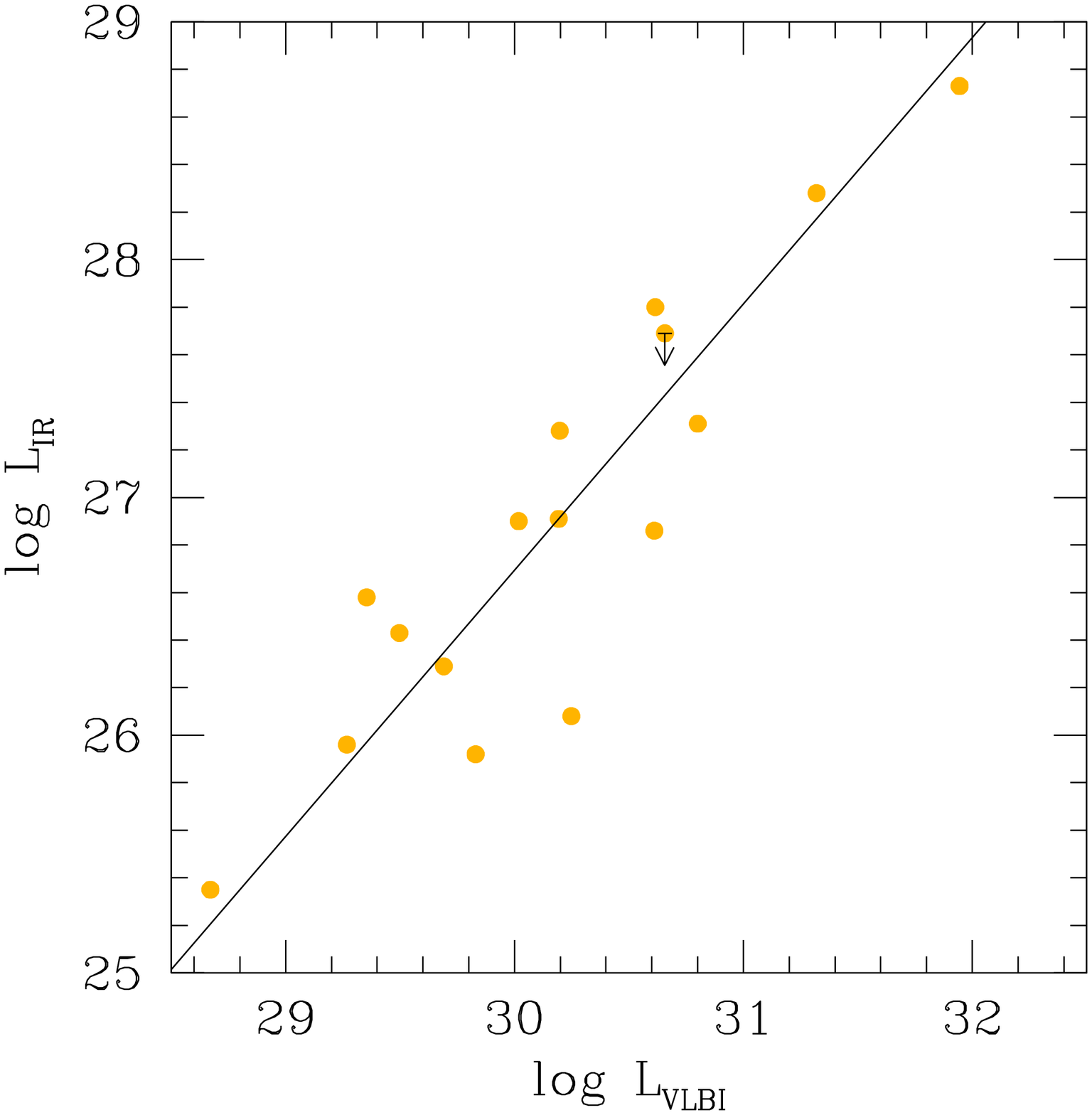}
\includegraphics[scale=0.45]{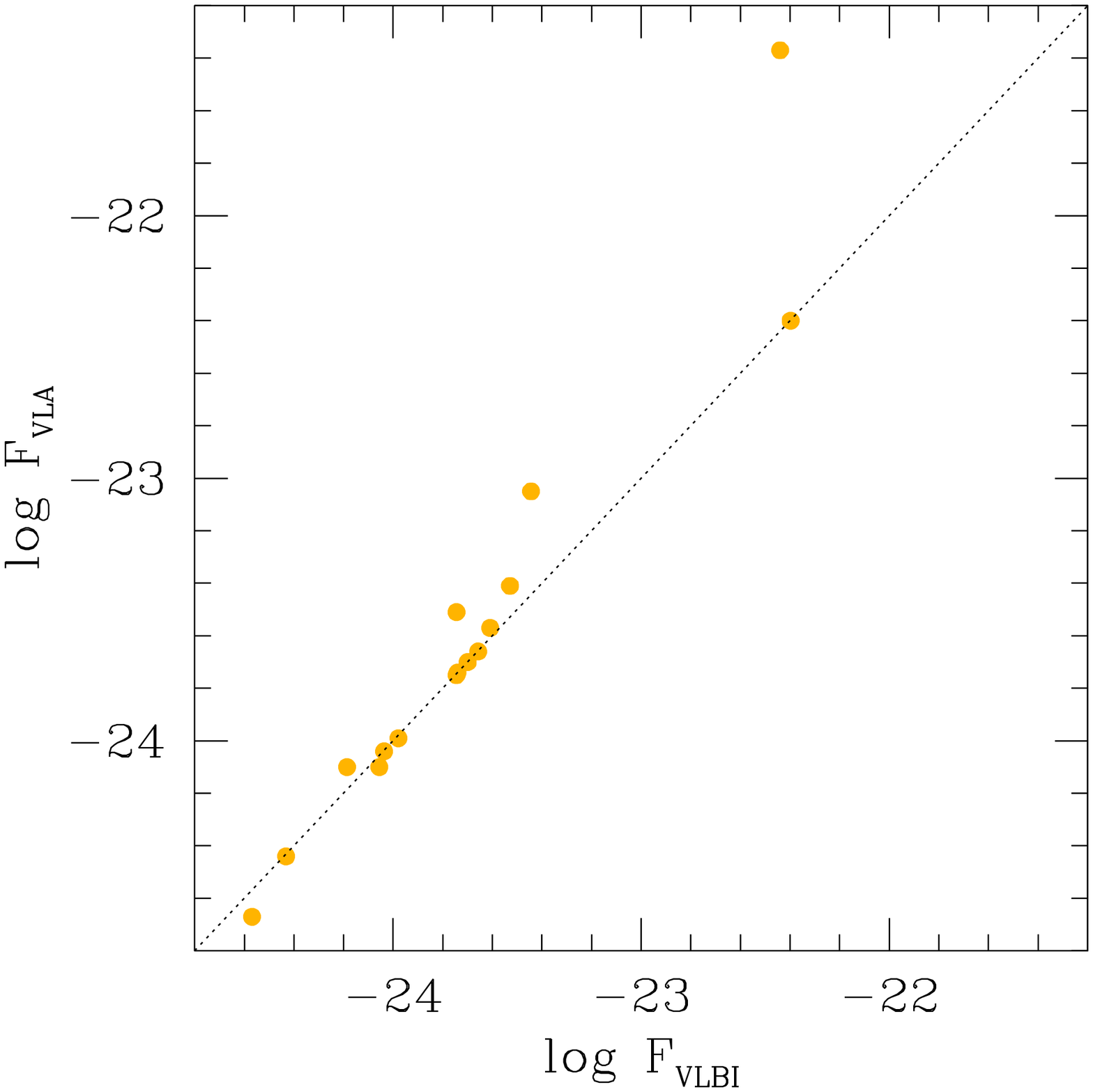}
\caption{VLBI data. In left panel we compare the VLBI radio and HST NIR 
nuclear luminosities (in erg s$^{-1}$ Hz$^{-1}$) for the FR~Is
  whose radio VLBI data are available. The solid lines represents the
  least-squared fits for the data shown in the plots. In right panel we
  compare the VLBI and VLA fluxes for the same FR~I sub-sample. The dotted
  lines represents the bisectrix of the plane.}
\label{vlbiir}
\end{figure*}

It is interesting to investigate the radio and NIR correlation for FR~Is by
considering radio data of higher angular resolution. To this aim, we take from
the literature the fluxes at 5 GHz obtained with VLBI observations.  The FR~Is
for which the VLBI radio data are available are listed in Table~5. In
Fig.~\ref{vlbiir} (left panel) we show the correlations in the
L$_{VLBI}$-L$_{IR}$ planes.  The linear correlation coefficient is r = 0.90 (P
=4.1$\times$10$^{-6}$).  These values as well as the slope and the intercept
of the derived correlations are essentially identical to those obtained with
the VLA data.  The reason of these similarities can be explained by looking at
Fig.~\ref{vlbiir} (right panel), which shows the VLA core flux of FR~Is
plotted vs. their VLBI flux. The figure clearly shows that the two
measurements lead to very similar flux values.  The dispersion is extremely
small, with the exception of only two outliers (3C~84 and NGC~6251), for which
the VLA flux is higher than that of VLBI by a factor of $\sim 2.5$ and $\sim
12$, respectively. However, since these two objects have a relatively high
core dominance, they are likely to be observed at a small angle with respect
to jet axis and thus they are more likely to be affected by flux variability.
That might explain the larger discrepancy between the VLA and VLBI fluxes for
these two sources with respect to the other FR~Is.  Summarizing, the
correlation between the VLBI and the IR nuclear luminosity is statistically
indistinguishable from that obtained using the VLA cores simply because the
core fluxes measured with VLA and VLBI are only marginally different.

\begin{table*}
\begin{center}
\caption{VLBI radio data}
\begin{tabular}{c|ccc}
\tableline\tableline
name            & F$_{VLBI}$ &  L$_{VLBI}$  & Ref \\
\tableline
      3C~31  &    -24.04   &      29.69 & G05 \\
      3C~66  &    -23.74   &      30.19 & G05 \\
    3C~83.1  &    -24.57   &      29.50 & X99 \\
      3C~84  &    -22.44   &      31.32 & K05 \\
     3C~264  &    -23.70   &      30.20 & G05 \\
     3C~270  &    -23.74   &      29.27 & J97 \\
   3C~272.1  &    -23.74   &      28.67 & G05 \\
     3C~274  &    -22.40   &      30.02 & G05 \\
     3C~296  &    -24.19   &      29.83 & G05 \\
     3C~310  &    -24.06   &      30.66 & G05 \\
     3C~317  &    -23.53   &      30.80 & V00 \\
     3C~338  &    -23.98   &      30.25 & G05 \\
     3C~346  &    -23.66   &      31.95 & G05 \\
     3C~449  &    -24.43   &      29.35 & G05 \\
     3C~465  &    -23.61   &      30.61 & G05 \\
\hline
   NGC~6251  &    -23.44   &      30.61 & J86 \\
\tableline
\end{tabular}
\tablecomments{Table of radio core flux and luminosity from VLBI
observations. Col. (1): name; Col. (2): radio core flux in erg s$^{-1}$
cm$^{-2}$ Hz$^{-1}$; Col. (3): radio core luminosity in erg s$^{-1}$
Hz$^{-1}$; Col. (4): References:G05: \citet{giovannini05}, X99: \citet{xu99a},
K05: \citet{kovalev05} (VLBA at 15GHz), J97: \citet{jones97} (VLBA at 1.6GHz),
V00: \citet{venturi00}, J86: \citet{jones86}.}
\end{center}
\end{table*}
\label{vlbidata}

\end{document}